\documentclass[a4paper,english,12pt]{article}
\usepackage[logonly]{trace}
\usepackage{babel}
\usepackage{amsfonts, amsmath, amssymb}
\usepackage{graphicx}
\usepackage{pstricks}
\usepackage{multirow}
\usepackage{fancyhdr}
\usepackage{vmargin, fancybox}
\usepackage[super,comma,sort&compress]{natbib}


\usepackage{times}
\newcommand{\dss}{\displaystyle}
\newcommand{\raro}{\rightarrow}
\newcommand{\be}{\begin{equation}}

\begin{document}

\renewcommand{\refname}{REFERENCES}

\large

\title{The interrelationship of integrable equations, differential
geometry and the geometry of their associated surfaces}

\author{Paul Bracken \\ Department of Mathematics \\ University of Texas \\
Edinburg, TX 78540 USA \footnote{bracken@panam.edu}} 

\date{January 2009}

\maketitle
\normalsize
\begin{abstract}
A survey of some recent and important results which have to
do with integrable equations and their relationship with the theory
of surfaces is given. Some new results are also presented.
The concept of the moving frame is examined, 
and it is used in several subjects which are discussed.
Structure equations are introduced in terms of differential forms.
Forms are shown to be very useful in relating geometry, equations and surfaces,
which appear in many sections. 
The topics of the chapters
are different and separate, but joined together by common themes
and ideas. Several subjects which are not easy to access
are reviewed and elaborated upon. These topics include Maurer-Cartan cocycles and recent
results with regard to generalizations of the Weierstrass-Enneper
system for generating constant mean curvature surfaces in three and
higher dimensional Euclidean spaces.
\end{abstract}

\vspace{2mm}
MSc: 53A05, 53A10, 53C15, 53C80, 53Z05, 35Q51
\vspace{2mm}

\newpage
\begin{center}
{\bf Table of Contents.}
\end{center}

\vspace{2cm}
\noindent
$1$. Introduction \dotfill\ 3-5

\noindent
$2.$ Classical Theory of Surfaces in Euclidean Three-Space \dotfill\ 5-7

\noindent
$3$. Surfaces on Lie Algebras, Lie Groups and Integrability \dotfill\ 7-18

\noindent
$4$. Differential Forms, Moving Frames and Surfaces \dotfill\ 18-28

$4.1$ Introduction to Moving Frames \dotfill\ 18

$4.2$ $SO (3)$ Lax Pair. \dotfill\  21

$4.3$ Nonlinear Partial Differential Equations Admitting $SO (2,1)$ Lax Pairs. \dotfill 26

\noindent
$5$. Maurer-Cartan Cocycles and Prolongations \dotfill\ 28-38

\noindent
$6$. The Generalized Weierstrass System Inducing Surfaces of
Constant and Nonconstant Mean Curvature \dotfill\ 39-51

$6.1$ Generalized Weierstrass Representations. \dotfill\ 39

$6.2$ A Physical Application Involving Nonlinear Sigma Models. \dotfill 47

$6.3$ Non-Constant Mean Curvature Surfaces.  \dotfill 48

\noindent
$7$. References. \dotfill 51-54
\newpage
\begin{center}
{\bf 1. INTRODUCTION}
\end{center}

The origins of soliton theory go back to the early part
of the nineteenth century, in particular, to the observation
of John Scott Russell in 1834 of a solitary bump-shaped
wave moving along a canal near Edinburgh. However, it wasn't until
1965 that this type of phenomena was rediscovered, in particular,
by Kruskal and Zabusky {\bf [1,2]} in the context of the
Fermi-Pasta-Ulam {\bf [3]} problem. It was they that coined the term soliton. 
In 1895, two Dutch mathematicians, Korteweg and de-Vries derived
a nonlinear wave equation which now bears their name.
It models long wave propagation in a rectangular channel and
has a traveling wave solution which resembles the solitary
wave observed by Russell. In fact, a pair of equations
equivalent to the KdV equation appeared even earlier in a work by
Boussinesq. It was not until the mid-twentieth century that 
the equation reappeared in work by such
researchers as Kruskal and Zabusky and Gardener and Morikawa in
1960 in an analysis of the transmission of hydromagnetic waves.
There continues to be ongoing interest in such nonlinear equations 
which arise in a diversity of systems such as the theory of
solids, liquids and gases {\bf [4,5]}. Self-localized nonlinear excitations 
are fundamental and inherent features of quasi-one-dimensional 
conducting polymers.
In 1962, Perring and Skyrme solved the sine-Gordon equation
numerically while using it in an elementary particle model.
The results generated from this equation were found not to disperse 
and two solitary waves were seen to keep their original shapes and velocities 
despite collisions.

In the pioneering work of Kruskal and Zabusky, the KdV 
equation was obtained as a continuum limit of an anharmonic
lattice model with cubic nonlinearity. The model displays the
existence of solitary waves. These waves have the remarkable 
property that they preserve both their amplitude and speed upon
interaction. These properties are the main reason for the use of the term soliton.
A soliton or solitary wave can be regarded as a solution
to any number of a variety of nonlinear partial differential equation.
In a more physical language, solitons have the
following striking properties. Energy is localized within a
small region and an elastic scattering phenomenon exists in the
interaction of two solitons. To put it another way, the shape
and velocity of the wave are recovered after an interaction
between such solutions. Solitons seem to behave as both
particle and wave. Originally they arose in the area of fluid
mechanics, and their study has extended into the areas of plasma
physics, nonlinear optics and classical and quantum field theory.
In large part, this is precisely due to the aforementioned 
properties. What is more, many branches of mathematics and
physics provide important tools for the study of solitons. It will
be seen here that the development of the study of solitons has 
resulted in reciprocal advances in many areas of mathematics as well.

Moreover, there is a deep connection between many of these
equations, the theory of surfaces {\bf [6]} and integrable systems {\bf [7,8]}.
It is the intention here to explore this interrelationship between
the theory of these equations and the surfaces that can be determined 
by them. It will be seen that many important ideas from the area
of differential geometry are applicable to the subjects studied here
and give the subject a unified perspective.

A generic method for the description of soliton interaction 
begins with a transformation which was originally introduced by
B\"acklund to generate pseudospherical surfaces.
Later Bianchi showed that the B\"acklund transformation admits
a commutativity property, a consequence of which is a nonlinear
superposition principle which is referred to as the
permutability theorem. As an example, both KdV and MKdV
equations reside in hierarchies which admit auto-B\"acklund 
transformations, nonlinear superposition principles as well as
multi-soliton solutions.

This article is a review of recent results by the author
as well as by other researchers. Let us begin with a brief
overview of the contents which follow. 
Although there are numerous ideas and themes which run throughout
the article, each chapter is separate and can be read on its own
independently of the others.
First, a review of
surface theory from the classical point of view will be
presented {\bf [6,9]}. The next section is a novel and active
area of interest which should appeal to those with an interest in this area.
The subject of the immersion of a two-dimensional surface into
a three-dimensional Euclidean space, as well as the $n$-dimensional
generalization, has been related to the problem of investigating
surfaces in Lie groups and in Lie algebras as well {\bf [10,11]}.
This gives an interesting correspondence between the Lax pair
of an integrable equation and their integrable surfaces. Using 
the formulation of the immersion of a two-dimensional surface
into three-dimensional Euclidean space, it will be shown that
a mapping from each symmetry of integrable equations to surfaces
in $\mathbb R^3$ can be established.

Next a differential forms approach to surfaces will be presented
{\bf [12-14]}. The problem of identifying whether a given nonlinear
partial differential equation admits a linear integrable system
is studied here by means of this differential geometric formalism {\bf [15]}.
It is shown that the fundamental equations of surface theory can be 
used to reproduce the compatibility conditions obtained from a linear
system in matrix form corresponding to a number of different Lie
algebras. In fact, the system of equations which has been obtained
from the linear matrix problem is derived from a system of 
differential forms and in combination with the first and second
fundamental forms leads to a link with surface theory in
differential geometry {\bf [16]}.

The subject of non-linear evolution equations and Maurer-Cartan
cocycles on $\mathbb R^2$ is introduced next. Maurer-Cartan cocycles
are defined and some general theoretical information about them is
provided. By using Maurer-Cartan cocycles and Cartan
prolongations for individual nonlinear equations, it is shown 
how B\"acklund transformations can be calculated for specific
equations. As an example, the B\"acklund transformation for the
sine-Gordon equation will be derived.

Finally, the subject of constant mean curvature surfaces has been
of great interest recently. From what has been discussed already,
the theory of surfaces has many applications in a great number
of areas of physical science. The theory of constant mean curvature 
surfaces has had a great impact on many problems which have physical
applications. In particular, there are many applications to such areas
as two-dimensional gravity, quantum field theory,
statistical physics and fluid dynamics {\bf [17,18]}. An application
of recent interest is the propagation of a string through space-time,
in which the particle describes a surface called its world sheet.
Thus, the subject of generalized Weierstrass representations,
in particular, the generalization of the Weierstrass-Enneper
approach due to B. Konopelchenko {\bf [19]} will be discussed
in detail. There exists a correspondence between this
representation and the two-dimensional nonlinear sigma model.
Both of these systems have been shown to be integrable, and
their symmetry groups have been calculated {\bf [20]}.
These symmetries have lead to the calculation in closed form
of many explicit solutions of the system, and the determination of
their soliton surfaces, as will be seen.

\begin{center}
{\bf 2. CLASSICAL THEORY OF SURFACES}
\end{center}

It is perhaps useful at this point to introduce some
classical results concerning surfaces, which arise out of
classical differential geometry. This will give a basic
review of surface theory and some preparation
for what is to follow.

Let ${\bf r} = {\bf r} (u,v)$ denote the position vector of
a generic point $P$ on a surface $\Sigma$ in $\mathbb R^3$.
Then the vectors ${\bf r}_u$ and ${\bf r}_v$ are tangential
to $\Sigma$ at $P$ and at such points at which they are
linearly independent, the quantity
$$
{\bf N} = \frac{{\bf r_u \times r_v}}{| {\bf r_u \times r_v}|},
\eqno(2.1)
$$
determines the unit normal to $\Sigma$.
The first and second fundamental forms of $\Sigma$ are defined by
$$
\begin{array}{c}
I = d s_I^2 = d {\bf r} \cdot d {\bf r}
= E \, du^2 + 2 F \, du dv + G \, dv^2,  \\
   \\
II = d s_{II}^2 =- d {\bf r} \cdot d {\bf N} =
e \, du^2 + 2 f du dv + g \, dv^2.  \\
\end{array}
\eqno(2.2)
$$
In (2.2), the coefficients are defined by
$$
\begin{array}{ccc}
E = {\bf r}_u \cdot {\bf r}_u,  &  F = {\bf r}_u \cdot {\bf r}_v, &  G = {\bf r}_v \cdot {\bf r}_v,  \\
  &    &   \\
e =- {\bf r}_u \cdot {\bf N}_u,  &  g= - {\bf r}_v \cdot {\bf N}_v, & f =- {\bf r}_u \cdot {\bf N}_v
=- {\bf r}_v \cdot {\bf N}_u.  \\
\end{array}
\eqno(2.3)
$$
There is a classical result of Bonnet which states that
$\{ E, F, G, e, f, g \}$ determines the surface $\Sigma$ up to its
position in space. The Gauss equations associated with $\Sigma$ 
are given as
$$
{\bf r}_{uu} = \Gamma^1_{11} {\bf r}_u + \Gamma^2_{11} {\bf r}_v+ e {\bf N},
\quad
{\bf r}_{uv} = \Gamma^1_{12} {\bf r}_u + \Gamma^2_{12} {\bf r}_v + f {\bf N},
\quad
{\bf r}_{vv} = \Gamma^1_{22} {\bf r}_u + \Gamma^2_{22} {\bf r}_v + g {\bf N},
\eqno(2.4)
$$
while the Weingarten equations are given by
$$
{\bf N}_u = \frac{f F - e G}{H^2} {\bf r}_u
+ \frac{e F - fE}{H^2} {\bf r}_v,
\quad
{\bf N}_v = \frac{gF - fG}{H^2} {\bf r}_u
+ \frac{fF - gE}{H^2} {\bf r}_v,
\eqno(2.5)
$$
where $H^2 = |{\bf r_u} \times {\bf r}_v|^2 = EG - F^2$. The
$\Gamma^i_{jk}$ are the Christoffel symbols and since the derivatives
of all the $\{ E, F, G, e, f, g \}$ with respect to $u$ and $v$
can be calculated from (2.3) and (2.4), the derivatives of all the
$\Gamma^i_{jk}$ can be calculated as well.

Thus, using these derivatives, the compatibility conditions
$( {\bf r}_{uu})_v = ( {\bf r}_{uv})_u$ and $({\bf r}_{uv})_v
= ( {\bf r}_{vv})_u$ applied to the linear Gauss system (2.4)
produces the nonlinear Mainardi-Codazzi system
$$
e_v- f_u = e \Gamma^1_{12} + f ( \Gamma^2_{12} - \Gamma^1_{11}) - g \Gamma^2_{11},
\qquad
f_v - g_u = e \Gamma^1_{22} + f ( \Gamma^2_{22} - \Gamma^1_{12}) - g \Gamma^2_{12}.
\eqno(2.6)
$$
The Theorema egregium of Gauss provides an expression for the
Gaussian or total curvature
$$
{\cal K} = \frac{eg - f^2}{EG - F^2},
\eqno(2.7)
$$
or in terms of $E$, $F$, and $G$ alone in Liouville's representation
$$
{\cal K} = \frac{1}{H} [ ( \frac{H}{E} \Gamma^2_{11})_v
- ( \frac{H}{E} \Gamma^2_{12})_u].
\eqno(2.8)
$$
If the total curvature of $\Sigma$ is negative, that is, if $\Sigma$ is
a hyperbolic surface, then the asymptotic lines on $\Sigma$ may be taken as
parametric curves. Then $e=g=0$ and the Mainardi-Codazzi equations
reduce to
$$
( \frac{f}{H} )_u + 2 \Gamma^2_{12} \frac{f}{H} =0,
\qquad
( \frac{f}{H} )_v + 2 \Gamma^1_{12} \frac{f}{H} =0.
\eqno(2.9)
$$
Moreover, we have
$$
{\cal K} =- \frac{f^2}{H^2} =- \frac{1}{\rho^2},
$$
$$
\Gamma^1_{12} = \frac{G E_v - F G_u}{2 H^2},
\qquad
\Gamma^2_{12} = \frac{E G_u - F E_v}{2 H^2}.
\eqno(2.10)
$$
The angle between the parametric lines is such that
$$
\cos \omega = \frac{F}{\sqrt{EG}},
\quad
\sin \omega = \frac{H}{\sqrt{EG}},
\eqno(2.11)
$$
and since $E$, $G>0$, we may take without loss of generality
$$
E= \rho^2 a^2, \quad
G = \rho^2 a^2, 
\quad
f = ab \rho \sin \omega.
\eqno(2.12)
$$
Then the Christoffel symbols are given by
$$
\Gamma^1_{12}= \frac{\rho_v a + \rho a_v - \cos \omega 
( \rho_u b + \rho b_u)}{\rho a \sin^2 \omega},
\qquad
\Gamma^2_{12} = \frac{\rho_u b + \rho b_u - \cos \omega
( \rho_v a + \rho a_v)}{\rho b \sin^2 \omega}.
\eqno(2.13)
$$
Substituting (2.13) into the pair (2.9), there results
$$
2 \rho a_v + 2 \rho_v a -2 \cos \omega ( \rho_u b + \rho b_u)
- \rho_va \sin^2 \omega =0,
\quad
2 \rho b_u + 2 \rho_u b - 2 \cos \omega ( \rho_v a + \rho a_v) 
- \rho_u b \sin^2 \omega =0.
\eqno(2.14)
$$
Solving the linear system in (2.14) for $a_v$ and $b_u$, we obtain 
$$
a_v + \frac{a}{2} \frac{\rho_v}{\rho} - \frac{1}{2} b
\frac{\rho_u}{\rho} \cos \omega =0,
\qquad
b_u + \frac{b}{2} \frac{\rho_u}{\rho} - \frac{1}{2} a
\frac{\rho_v}{\rho} \cos \omega =0.
\eqno(2.15)
$$
The representation for the total curvature is
$$
\omega_{uv} + \frac{1}{2} ( \frac{\rho_u}{\rho} \frac{b}{a}
\sin \omega )_u + \frac{1}{2} ( \frac{\rho_v}{\rho} \frac{a}{b}
\sin \omega )_v - ab \sin \omega =0.
\eqno(2.16)
$$
For the particular case in which ${\cal K} =- 1/ \rho^2 <0$ is a
constant, $\Sigma$ is referred to as a pseudospherical surface.
Then (2.15) implies that $a = a(u)$, $b= b(v)$, and if
$\Sigma$ is now parametrized by arc length along asymptotic lines,
then
$$
d s_I^2 = du^2 + 2 \cos \omega \, du dv + dv^2,
\qquad
d s_{II}^2 = \frac{2}{\rho} \sin \omega \, du dv.
\eqno(2.17)
$$
Equation (2.16) then reduces to the sine-Gordon equation
$$
\omega_{uv} = \frac{1}{\rho^2} \sin \omega.
\eqno(2.18)
$$
Thus, there is a clear indication of a relationship between
surfaces and an integrable equation.

\begin{center}
{\bf 3. SURFACES ON LIE ALGEBRAS, LIE GROUPS AND INTEGRABILITY}
\end{center}

There have been some interesting developments recently related
to the problem of the immersion of a 2-dimensional surface into a 
3-dimensional Euclidean space, as well as the $n$-dimensional
generalization {\bf [21,22,23]}. These will be reviewed here.
This subject has been shown to be related to the problem of
studying surfaces in Lie groups and Lie algebras {\bf [24]}. 
It has been found useful for investigating integrable surfaces,
or surfaces which are described by integrable equations.
Starting from a suitable Lax pair, which implies a suitable
integrable equation, it is possible to construct explicitly
large classes of integrable surfaces.

Let ${\bf F} = ( F_1, F_2, F_3) : \pi \rightarrow \mathbb R^3$
be an immersion of a domain $\pi \subset \mathbb R^2$ into
3-dimensional Euclidean space. For $(u,v) \in \pi$, the
Euclidean metric induces a metric with coefficients 
$g_{ij} (u,v)$ on the surface. These functions and $d_{ij} (u,v)$,
which define the second fundamental form, satisfy a system
of three nonlinear equations known as the Gauss-Codazzi equations,
which are the compatibility condition of the Gauss-Weingarten
system. There exist two geometrical characteristics on such a
surface known as the Gauss curvature $K$ and the mean curvature
$H$. Some results will be given in other sections which correspond 
to constant $K$ and constant $H$.

A surface will be called integrable if and only if its Gauss-Codazzi
equations are integrable. Integrable equations also arise as the
compatibility condition of a pair of linear equations, which is
usually referred to as a Lax pair. Here we want to show this problem
is closely related to the problem of studying surfaces in Lie groups 
and Lie algebras.

Let $G$ be a group and ${\cal G}$ the Lie algebra of $G$. Assume
there exists an invariant scalar product in ${\cal G}$. The scalar
product will not be degenerate so there exists an orthonormal basis
$\{ e_i \}$ in ${ \cal G }$ such that $\langle e_i, e_j \rangle =
\delta_{ij}$. To introduce a surface in $G$, let $\Phi (u,v) \in G$
for every $(u,v)$ in some neighborhood of $\mathbb R^2$.
There exists a canonical map from the tangent space of $G$ to the
Lie algebra ${\cal G}$. If $\Phi_u$ and $\Phi_v$ are the tangent
vectors of $\Phi$ at the point $(u,v)$, this map is defined by
$$
\frac{\partial \Phi}{\partial u} \Phi^{-1} = U_j e_j,
\qquad
\frac{\partial \Phi}{\partial v} \Phi^{-1} = V_{j} e_j,
\eqno(3.1)
$$
where $U_j$ and $V_j$ are some functions of $(u,v)$ and
$j=1, \cdots, n$. Equations (3.1) define $\Phi$ through its
value in the Lie algebra. Suppose the structure constants of
${\cal G}$ satisfy
$$
[ e_k , e_m ] = c_{kmj} e_j,
\eqno(3.2)
$$
with summation implied. Differentiating the first equation
of (3.1) with respect to $v$ and the second with respect to $u$,
then upon subtracting these we have
$$
\frac{\partial U_j}{\partial v} e_j \Phi + U_j e_j \frac{\partial \Phi}
{\partial v} - \frac{\partial V_j}{\partial u} e_j \Phi
- V_j e_j \frac{\partial \Phi}{\partial  u}
= ( \frac{\partial U_j}{\partial v} - \frac{\partial V_j}{\partial u}) e_j \Phi
+ ( U_m V_s e_m e_s - V_s U_m e_s e_m ) \Phi =0.
\eqno(3.3)
$$
Expression (3.3) implies that
$$
( \frac{\partial U_j}{\partial v} -
\frac{\partial V_j}{\partial u} ) e_j + U_m V_s c_{msj} e_j =0.
$$
which when written just in terms of $U$ and $V$, this is written
$$
\frac{\partial U}{\partial v}
- \frac{\partial V}{\partial u} + [ U, V] =0.
\eqno(3.4)
$$
This result can be summarized next.

{\em Proposition 3.1.} Let $\Phi (u,v) \in G$ be a differentiable
function of $u$, $v$ for every $(u,v)$ in some neighborhood of
$\mathbb R^2$. Then $\Phi$ defined by (3.1) exists if and
only if the functions $U_j$ and $V_j$ satisfy (3.4).

To introduce a surface in ${\cal G}$, let $F(u,v) \in {\cal G}$ 
for every $(u,v)$ in a neighborhood of $\mathbb R^2$. The first
fundamental form of $F$ is defined by
$$
\langle \frac{\partial F}{\partial u},
\frac{\partial F}{\partial u} \rangle \, du^2
+ 2 \langle \frac{\partial F}{\partial u}, 
\frac{\partial F}{\partial v} \rangle \, du dv
+ \langle \frac{\partial F}{\partial v},
\frac{\partial F}{\partial v} \rangle \, dv^2.
\eqno(3.5)
$$
Let $N^{(s)} \in {\cal G}$, $s=1, \cdots, n-2$ be the
elements of ${\cal G}$ defined by $\langle N^{(l)}, N^{(l)} \rangle=1$,
$\langle F_u, N^{(l)} \rangle = \langle F_v, N^{(l)} \rangle =0$.
Then the second fundamental forms of $F$ are defined by
$$
\langle \frac{\partial^2 F}{\partial u^2}, N^{(s)} \rangle \, du^2
+ 2 \langle \frac{\partial^2 F}{\partial u \partial v},
N^{(s)} \rangle \, du dv
+ \langle \frac{\partial^2 F}{\partial v^2}, N^{(s)} \rangle \, dv^2,
\eqno(3.6)
$$
for $s=1, \cdots, n-2$. Surfaces in $G$ can be related to
surfaces in ${\cal G}$ by using the adjoint representation to
write
$$
\frac{\partial F}{\partial u} = \Phi^{-1} a_j e_j \Phi,
\qquad
\frac{\partial F}{\partial v} = \Phi^{-1} b_j e_j \Phi,
\eqno(3.7)
$$
where $a_j$ and $b_j$ are some functions of $(u, v)$.
Differentiating the first expression in (3.7) with
respect to $v$ and using the fact that 
$(\Phi^{-1})_{\tau} =- \Phi^{-1} \Phi_{\tau} \Phi^{-1}$, then
modulo (3.7), we obtain
$$
\frac{\partial^2 F}{\partial u \partial v} =- \Phi^{-1} V_s e_s a_j e_j \Phi
+ \Phi^{-1} \frac{\partial a_j}{\partial v} e_j \Phi
+ \Phi^{-1} a_j e_j V_s e_s \Phi
= \Phi^{-1} ( \frac{\partial a_j}{\partial v} e_j - V_sa_m c_{smj} e_j ) \Phi.
\eqno(3.8)
$$
In a similar way, differentiating $F_v$ with respect to $u$,
we have
$$
\frac{\partial^2 F}{\partial u \partial v} = \Phi^{-1}
( \frac{\partial b_j}{\partial u} e_j - U_s b_m c_{smj} e_j ) \Phi.
\eqno(3.9)
$$
Requiring that the derivatives in (3.8) and (3.9) match
gives the following result.

{\em Proposition 3.2.} Let $\Phi (u,v) \in G$ be a surface defined by (3.1).
Let $F (u,v) \in {\cal G}$ be a differentiable function of $u$ and
$v$ for every $(u,v)$ in some neighborhood of $\mathbb R^2$. Then (3.7)
defines a surface $F (u,v) \in {\cal G}$ if and only if $a_j$ and
$b_j$ satisfy
$$
\frac{\partial a_j}{\partial v} + a_k V_m c_{kmj} =
\frac{\partial b_j}{\partial u} + b_k V_m c_{kmj},
\quad
k,m,j =1, \cdots, n.
$$

It is often possible to calculate $a_j$, $b_j$ and $F$
explicitly.

{\em Theorem 3.1.} Let $U_j (u,v)$ and $V_j (u,v)$, 
$j=1, \cdots, n$ be differentiable functions of $u$ and
$v$ for every $(u,v)$ in some neighborhood of $\mathbb R^2$.
Let $\{ e_j \}_{j=1}^n$ be an orthonormal basis in the
Lie algebra ${\cal G}$ of the Lie group $G$.

Suppose that $U_j$ and $V_j$ depend on a parameter
$\lambda$ and satisfy (3.3), where $c_{kmj}$, $k$, $m$,
$j=1, \cdots, n$ are the structure constants associated
with ${\cal G}$, but $\lambda$ does not appear
explicitly in (3.4). Define $U$ and $V$ as follows
$$
U = U_j e_j,
\qquad
V = V_j e_j.
\eqno(3.10)
$$

$(i)$ If $A$ and $B$ are defined to be
$$
\begin{array}{c}
A = a_j e_j = \alpha_1 \dss \frac{\partial U}{\partial u}
+ \alpha_2 \frac{\partial U}{\partial v}
+ \alpha_3 \frac{\partial U}{\partial \lambda}
+ \alpha_4 \frac{\partial}{\partial u} (u U)
+ \alpha_5 v \frac{\partial U}{\partial v} + [U, M],   \\
  \\
B= b_j e_j = \alpha_1 \dss \frac{\partial V}{\partial u}
+ \alpha_2 \frac{\partial V}{\partial v}
+ \alpha_3 \frac{\partial V}{\partial \lambda}
+ \alpha_4 u \frac{\partial V}{\partial u}
+ \alpha_5 \frac{\partial}{\partial v} (v V) + [ V, M],  \\
\end{array}
\eqno(3.11)
$$
where $M = m_j e_j$ and $ \alpha_1, \cdots, \alpha_5,
m_1, \cdots, m_n$ are constant scalars, then the equations
$$
\frac{\partial F}{\partial u} = \Phi^{-1} A \Phi,
\qquad
\frac{\partial F}{\partial v} = \Phi^{-1} B \Phi,
\eqno(3.12)
$$
are compatible, and can be used to define a surface 
$F (u,v) \in {\cal G}$.

$(ii)$ The solution of (3.12) where $A$ and $B$ are defined
by (3.11) is, to within an additive constant, given by
$$
F = \alpha_1 \Phi^{-1} U \Phi + \alpha_2 \Phi^{-1} V \Phi
+ \alpha_3 \Phi^{-1} \frac{\partial \Phi}{\partial \lambda}
+ \alpha_4 u \Phi^{-1} U \Phi + \alpha_5 v \Phi^{-1} V \Phi
- \Phi^{-1} M \Phi.
\eqno(3.13)
$$

{\em Proof:} The equations (3.1) are compatible if and only if 
(3.4) holds. From the equations for $F$, we determine that
$$
\frac{\partial^2 F}{\partial v \partial u} = \frac{\partial \Phi^{-1}}
{\partial v} A \Phi + \Phi^{-1} \frac{\partial A}{\partial v} \Phi
+ \Phi^{-1} A \frac{\partial \Phi}{\partial v}
= - \Phi^{-1} V_m e_m A \Phi + \Phi^{-1} \frac{\partial A}{\partial v} \Phi
+ \Phi^{-1} A V_m e_m \Phi
$$
$$
= \Phi^{-1} ( \frac{\partial A}{\partial v} - [ V,A] ) \Phi.
\eqno(3.14)
$$
Similarly,
$$
\frac{\partial^2 F}{\partial u \partial v} = \Phi^{-1}
( \frac{\partial B}{\partial u} + [ B, U] ) \Phi.
\eqno(3.15)
$$
Upon equating the derivatives in (3.14) and (3.15)
and moving all terms to the same side, it follows that
$$
\frac{\partial A}{\partial v} - \frac{\partial B}{\partial u}
+ [ A, V] + [ U, B] =0.
\eqno(3.16)
$$
Suppose $A$ and $B$ are defined by (3.11) and $U$ and $V$
satisfy (3.4), then by direct calculation, we have
$$
\frac{\partial A}{\partial v} - \frac{\partial B}{\partial u}
= \alpha_1 \frac{\partial}{\partial u} 
( \frac{\partial U}{\partial v} - \frac{\partial V}{\partial u})
+ \alpha_2 \frac{\partial}{\partial v}
( \frac{\partial U}{\partial v} - \frac{\partial V}{\partial u})
+ \alpha_3 ( \frac{\partial^2 U}{\partial v \partial \lambda}
- \frac{\partial^2 V}{\partial u \partial \lambda})
+ \alpha_4 ( \frac{\partial}{\partial v} (u U) - u
\frac{\partial V}{\partial u})
$$
$$
+ \alpha_5 \frac{\partial}{\partial v} ( v \frac{\partial U}{\partial v}
- \frac{\partial}{\partial u} (v V)) +
[ \frac{\partial U}{\partial v}, M] - [ \frac{\partial V}{\partial u},M]
$$
$$
= \alpha_1 \frac{\partial}{\partial u} [ U, V] - \alpha_2
\frac{\partial}{\partial v} [ U, V] - \alpha_3
\frac{\partial}{\partial \lambda} [U, V]
- \alpha_4 \frac{\partial}{\partial u} (u [M, V] )
- \alpha_5 \frac{\partial}{\partial v} ( v [ U, V])
\eqno(3.17)
$$
$$
+ [ \frac{\partial U}{\partial v} - \frac{\partial V}{\partial u}, M].
$$
Using $A$ and $B$ given in (3.11), let us work out the terms
of $[A, V] + [U, B]$ according to each coefficient $\alpha_j$ one at a time,
$$
\alpha_1 [ \frac{\partial U}{\partial u}, V] + \alpha_1 
[ U, \frac{\partial V}{\partial u} ] = \alpha_1 \frac{\partial}{\partial u}
[U, V],
$$
$$
\alpha_2 [ \frac{\partial U}{\partial v}, V] + \alpha_2 [ U,
\frac{\partial V}{\partial v} ]
= \alpha_2 \frac{\partial}{\partial v} [ U, V],
$$
$$
\alpha_3 [ \frac{\partial U}{\partial \lambda}, V] 
+ \alpha_3 [U, \frac{\partial V}{\partial \lambda} ]
= \alpha_3 \frac{\partial}{\partial \lambda} [U,V],
\eqno(3.18)
$$
$$
\alpha_4 [ \frac{\partial}{\partial u} (u U), V] + \alpha_4 [ U, 
u \frac{\partial V}{\partial u}] = \alpha_4
\frac{\partial}{\partial u} ( u [U,V]),
$$
$$
\alpha_5 [ v \frac{\partial U}{\partial v}, V]
+ \alpha_5 [U, \frac{\partial}{\partial v} (vV)]
= \alpha_5 \frac{\partial}{\partial v} (v [U,V ]).
$$
Finally, using Jacobi's identity, we can write
$$
[[U,M],V] + [U, [V,M]]=
[V, [M,U]]+[U,[V,M]]
= [[U,V],M] =- [ \frac{\partial U}{\partial v} -
\frac{\partial V}{\partial u}, M].
$$
Substituting all of these results for the brackets (3.18)
as well as for $A_v-B_u$ from (3.17) into the left-hand
side of (3.16), it can be seen that (3.16) is 
satisfied identically.

$(ii)$ To prove that $F$ given by (3.13) satisfies (3.12),
differentiate $F$ with respect to $u$ to obtain
$$
\frac{\partial F}{\partial u} =- \alpha_1 \Phi^{-1}UU \Phi
+ \alpha_1 \Phi^{-1} \frac{\partial U}{\partial u} \Phi
+ \alpha_1 \Phi^{-1} UU \Phi - \alpha_2 \Phi^{-1} UV \Phi
+ \alpha_2 \Phi^{-1} \frac{\partial V}{\partial u} \Phi
+ \alpha_2 \Phi^{-1} VU \Phi
$$
$$
- \alpha_3 \Phi^{-1} U \frac{\partial \Phi}{\partial \lambda} 
+ \alpha_3 \Phi^{-1} \frac{\partial^2 \Phi}{\partial u \partial \lambda}
+ \alpha_4 \Phi^{-1} U \Phi - \alpha_4 u \Phi^{-1} UU \Phi
+ \alpha_4 u \Phi^{-1} \frac{\partial U}{\partial u} \Phi
+ \alpha_4 u \Phi^{-1} UU \Phi
$$
$$
- \alpha_5 v \Phi^{-1} UV \Phi + \alpha_5 v \Phi^{-1}
\frac{\partial V}{\partial u} \Phi + \alpha_5 v \Phi^{-1} VU \Phi
+ \Phi^{-1} UM \Phi - \Phi^{-1}MU \Phi
$$
$$
= \Phi^{-1} \{  \alpha_1 \frac{\partial U}{\partial u} +
\alpha_2 ( \frac{\partial V}{\partial u} - [ U, V] )
- \alpha_3 \frac{\partial U}{\partial \lambda}
+ \alpha_4 \frac{\partial}{\partial u} (u U)
+ \alpha_5 ( v \frac{\partial V}{\partial u} - v [U, V] )
+ [U, M] \}  \Phi.
$$
Using (3.4), this simplifies to the form
$$
\frac{\partial F}{\partial u} = \Phi^{-1}
\{ \alpha_1 \frac{\partial U}{\partial u} + \alpha_2
\frac{\partial V}{\partial v}
- \alpha_3 \frac{\partial U}{\partial \lambda}
+ \alpha_4 \frac{\partial}{\partial u} (uU)
+ \alpha_5 v \frac{\partial U}{\partial v}
+ [U, M] \} \Phi = \Phi^{-1} A \Phi,
$$
as required. Similarly, the derivative of $F$ with respect
to $v$ is calculated in the same way, and the second equation
of (3.12) then results.

Using a variation of parameter, it follows that this $F$ is
unique to within a constant matrix. $\clubsuit$

This is really a consequence of the fact that (3.16) is the
variational equation of (3.4). In fact, if $U$ and $V$ are replaced
by $U + \epsilon A$ and $V + \epsilon B$, then the $O( \epsilon)$ term
of (3.4) yields (3.16). This means that every symmetry of (3.4)
implies a solution of (3.16).

It will be useful and instructive to write down the previous Theorem
for the case in which the group $G$ is $SU (2)$. In this case,
$e_j =-i \sigma_j$, for $j=1,2,3$ where $\sigma_j$ are the
Pauli matrices given by
$$
\sigma_1 = \left(
\begin{array}{cc}
0  &  1  \\
1  &  0  \\
\end{array}  \right),
\qquad
\sigma_2 = \left(
\begin{array}{cc}
0  &  -i  \\
i  &   0   \\
\end{array}  \right),
\qquad
\sigma_3 =  \left(
\begin{array}{cc}
1  &  0   \\
0  &  -1  \\
\end{array}  \right),
\eqno(3.19)
$$
and the structure constants are given by $c_{ijk} = 2 \epsilon_{ijk}$,
where $\epsilon_{ijk}$, $i,j,k =1,2,3$ is the usual
antisymmetric tensor.

To the vector ${\bf F} = (F_1, F_2, F_3 )^T \in \mathbb R^3$,
we associate the matrix ${\bf F} = F_j e_j \in su (2)$, which we write
$$
{\bf F} =- i F_j e_j = \left(
\begin{array}{cc}
-i F_3  & -F_2 -i F_1  \\
F_2  -i F_1  &  i F_3  \\
\end{array}  \right).
\eqno(3.20)
$$
The problem of immersing the 2-dimensional surface $x_j = F_j (u,v)$,
$j=1,2,3$ into 3-dimensional space becomes the problem of studying the 
relationship between the 3-dimensional sphere $\Phi (u,v) \in SU (2)$
and the two-dimensional surface $F (u,v) \in su (2)$. Thus taking
$e_j =-i \sigma_j$ and $ c_{ijk} = 2 \epsilon_{ijk}$ in Theorem 3.1,
this theorem can be restated for the case of $SU (2)$.

{\em Theorem 3.2.} Let $U (u,v)$ and $V (u,v) \in su (2)$ be differentiable 
functions of $u$ and $v$ for every $(u,v)$ in some neighborhood of
$\mathbb R^2$. Assume that the functions $U$ and $V$ satisfy 
equation (3.4). Then the equations
$$
\frac{\partial \Phi}{\partial u} = U \Phi,
\qquad
\frac{\partial \Phi}{\partial v} = V \Phi,
\eqno(3.21)
$$
define a 2-dimensional surface $\Phi (u,v) \in SU (2)$.

Let $A (u,v)$ and $B (u,v) \in su (2)$ be real, differentiable
functions of $u$ and $v$ for every $(u,v)$ in some neighborhood
of $\mathbb R^2$. In addition to this, assume that these 
functions satisfy (3.16). Then equations (3.12) together with
${\bf F} =- i F_j \sigma_j$ define a 2-dimensional surface
$x_j = F_j (u,v) \in \mathbb R$, $j=1,2,3$ in a 3-dimensional
Euclidean space. The first and second fundamental forms of
this surface are
$$
\langle A, A \rangle \, du^2 + 2 \langle A, B \rangle \, du dv
+ \langle B, B \rangle \, dv^2,
\eqno(3.22)
$$
and
$$
\langle \frac{\partial A}{\partial u} + [A,U], C \rangle \, du^2
+ 2 \langle \frac{\partial A}{\partial v} + [ A,V ], C \rangle \, du dv
+ \langle \frac{\partial B}{\partial v} + [ B,V], C \rangle \, dv^2,
\eqno(3.23)
$$
respectively. In (3.22) and (3.23), we have
$$
\langle A, B \rangle =- \frac{1}{2} tr \, (A, B),
\qquad
C = \frac{[A, B ] }{| [A, B]|},
\qquad
|A| = \sqrt{ \langle A, A \rangle }.
\eqno(3.24)
$$
$\clubsuit$

Let us consider the following example. In Theorem 3.2, let us put
$A = -i a \sigma_1$, $B =- i ( b_1 \sigma_1 + b_2 \sigma_2 )$,
$U =- \frac{i}{2} U_j \sigma_j$ and $V =- \frac{i}{2} V_j \sigma_j$
into (3.16). After using the commutation relations, we obtain
$$
( - \frac{\partial a}{\partial v} + \frac{\partial b_1}{\partial u} 
+ b_2 U_3 ) \sigma_1 + ( \frac{\partial b_2}{\partial u} + a V_3 - b_1 U_3 )
\sigma_2 +( -a V_2 + b_1 U_2 - b_2 U_1 ) \sigma_3 =0.
\eqno(3.25)
$$
For this to hold, the coefficients of $\sigma_j$ must vanish
giving the system of equations
$$
\frac{\partial a}{\partial v} - \frac{\partial b_1}{\partial u} - b_2 U_3 =0,
\qquad
\frac{\partial b_2}{\partial u} + a V_3 - b_1 U_3 =0,
\qquad
a V_2 - b_1 U_2 + b_2 U_1 =0.
\eqno(3.26)
$$
The first and third equations of (3.26) imply that
$$
U_3 = \frac{1}{b_2} ( \frac{\partial a}{\partial v} 
- \frac{\partial b_1}{\partial u}),
\qquad
V_2 = \frac{1}{a} [ b_1 U_2 - b_2 U_1 ].
$$
Putting these results in the second equation of (3.26) gives
$$
V_3 = \frac{1}{a b_2} [ b_1 \frac{\partial a}{\partial v}
- b_1 \frac{\partial b_1}{\partial u} - b_2
\frac{\partial b_2}{\partial u} ].
$$
Let $\Phi \in SU (2)$, then the equations for $F$ are
obtained by substituting $A$ and $B$. They take the form
$$
\frac{\partial F}{\partial u} =- \Phi^{-1} a \sigma_1 \Phi,
\qquad
\frac{\partial F}{\partial v} = -i \Phi^{-1}
( b_1 \sigma_1 + b_2 \sigma_2 ) \Phi.
\eqno(3.27)
$$
Then $F$ defines a 2-dimensional surface in $\mathbb R^3$.
Moreover,
$$
\langle A, A \rangle = a^2,
\qquad
\langle A, B \rangle = a b_1,
\qquad
\langle B,B \rangle = b_1^2 + b_2^2,
$$
give the coefficients of the first fundamental form (3.22),
which can be written
$$
I = a^2 \, du^2 + 2 a b_1 \, du dv + ( b_1^2 + b_2^2 ) \, dv^2.
\eqno(3.28)
$$
With $A =- i a \sigma_1$ and $B =- i ( b_1 \sigma_1
+ b_2 \sigma_2)$. we calculate $[ A,B ] =- [a \sigma_1,
b_1 \sigma_1 + b_2 \sigma_2 ]
=-2 a b_2 \sigma_3$ and $ C=- \sigma_3$. Therefore,
$[A, U ] =- \frac{1}{2} a U_j [ \sigma_1, \sigma_j]
= -a U_2 \sigma_3$ and
$$
\langle \frac{\partial A}{\partial u} + [A, U] , C \rangle
= \langle -i \frac{\partial a}{\partial u} \sigma_1 - a U_2 \sigma_3,
- \sigma_3 \rangle = a U_2.
$$
Similarly, it follows that $[ A, V] =- a ( \sigma_3 V_2
- \sigma_2 V_3)$, and we must have
$$
\langle \frac{\partial A}{\partial v} + [ A, V], C \rangle = a V_2,
\qquad
\langle \frac{\partial B}{\partial v} + [ B, V], C \rangle
= b_1 V_2 - b_2 V_1.
$$
Putting these together in (3.23), the second fundamental form
is given by
$$
II = a U_2 \, du^2 + 2 a V_2 \, du dv + ( b_1 V_2 - b_2 V_1) \, dv^2.
\eqno(3.29)
$$
In terms of matrices, these fundamental forms are given by
$$
I = \left(
\begin{array}{cc}
a^2  & a b_1  \\
a b_1  &  b_1^2 + b_2^2  \\
\end{array}   \right),
\qquad
II =  \left(
\begin{array}{cc}
a U_2  &   a V_2  \\
a V_2  &  b_1 V_2 - b_2 V_1  \\
\end{array}   \right).
\eqno(3.30)
$$
The Gauss and mean curvature are defined by
$$
K = \det (II \cdot I^{-1}) =- ( \frac{U_1}{a})^2 + \frac{U_2}{a}
( \frac{b_1 U_1 - a V_1}{a b_2} ),
\qquad
H =- tr (II \cdot I^{-1})
= - \frac{U_2}{a} - \frac{b_1 U_1 -a V_1}{a b_2}.
\eqno(3.31)
$$
It can be shown the surface $F$ is unique up to position
in space. Given the fundamental forms (3.30), $U_2$, $V_2$ and $V_1$
can be solved for. Since these functions satisfy the Gauss-Codazzi
equations (3.4), $\Phi \in SU (2)$ can be defined by (3.21) to
within three constants. Equations (3.12) imply $F \in su (2)$ within 
three additional constants.These six arbitrary constants
correspond to arbitrary motions of the surface in $\mathbb R^3$. Indeed, 
the transformations $\hat{F}= f F f^{-1} + \tilde{A}$, $\hat{N} =
f N f^{-1}$, $\hat{\Phi} = \Phi f$, $f \in SU (2)$, $\tilde{A} \in su (2)$
leave (3.21) and the fundamental forms invariant. The constants of
$\tilde{A}$ introduce a translation while the constants
of $f$ introduce a rotation. Therefore, six arbitrary constants appear.

Let us summarize this collection of results in Theorem 3.3.

{\em Theorem 3.3.} Let $U_1$, $U_2$, $V_1$, $a$, $b_1$ and
$b_2$ such that $a \neq 0$, $b_2 \neq 0$ be real
differentiable functions of $u$ and $v$ for every $(u,v)$ in
some neighborhood of $\mathbb R^2$. Assume that these functions
satisfy the Gauss-Codazzi equations (3.4), where $U_3$, $V_2$ and $V_3$
are defined by
$$
U_3 = \frac{1}{b_2} ( \frac{\partial a}{\partial v} - 
\frac{\partial b_1}{\partial u}),
\qquad
V_2 = \frac{1}{a} ( b_1 U_2 - b_2 U_1),
\qquad
V_3 = \frac{1}{a b_2} ( b_1 \frac{\partial a}{\partial v}
- b_1 \frac{\partial b_1}{\partial u}
- b_2 \frac{\partial b_2}{\partial u} ).
\eqno(3.32)
$$
Let $\Phi \in SU(2)$ be defined by (3.21). Then the equations
$$
\frac{\partial F}{\partial u} =-i \Phi^{-1} a \sigma_1 \Phi,
\qquad
\frac{\partial F}{\partial v} =- i \Phi^{-1}
( b_1 \sigma_1 + b_2 \sigma_2 ) \Phi,
\eqno(3.33)
$$
where $\sigma_j$ are the Pauli matrices (3.19), define a
2-dimensional surface
$$
x_j = F_j (u,v),
\qquad
j=1,2,2,
\eqno(3.34)
$$
in $\mathbb R^3$. Its first and second fundamental forms
are given in (3.28) and (3.29). The Gauss and mean curvatures
are given in (3.31). This surface is unique to within
position in space.

To close this section, a final result and an application along
these lines is presented below and gives an explicit construction
of functions $A$ and $B$ as well as the immersion function $F$
based on the symmetries of (3.4) and (3.1) {\bf [25]}. 

{\em Theorem 3.4.} Suppose that $U (u,v)$ and $V (u,v)$ can be
parametrized in terms of $\lambda$ and a scalar function
$\theta (u,v)$ in such a way that (3.4) is equivalent to a
single partial differential equation for $\theta (u,v)$ 
independent of $\Lambda$. This equation, which by definition
is called an integrable partial differential equation, possesses
the Lax pair defined by (3.21). Define the $su (2)$ valued functions
$ A (u,v, \lambda)$ and $B (u,v, \lambda)$ by
$$
A = \alpha \frac{\partial U}{\partial \lambda}
+ \frac{\partial M}{\partial u} + [ M,U] + U' \phi,
\eqno(3.35)
$$
$$
B = \alpha \frac{\partial V}{\partial \lambda} 
+ \frac{\partial M}{\partial v} + [M,V] + V' \phi,
\eqno(3.36)
$$
where $\alpha (\lambda)$ is an arbitrary scalar function
of $\lambda$. Also, $M (u,v; \lambda)$ is an $su (2)$ valued
arbitrary function of $u$, $v$ and $\lambda$ and the scalar 
$\phi$ is a symmetry of the partial differential equation
satisfied by the function $\theta (u,v)$. The prime denotes Fr\'echet
differentiation. Then there exists a surface with immersion
$F (u,v; \lambda)$ defined in terms of $A$, $B$, $\Phi$ by 
(3.35) and (3.36). Furthermore, $F$ to within an additive
constant, is given by
$$
F= \Phi^{-1} ( \alpha \frac{\partial \Phi}{\partial \lambda}
+ M \Phi + \Phi' \phi).
\eqno(3.37)
$$

{\bf Proof:} This is similar to Theorem 3.2, so we just
verify (3.16),
$$
\frac{\partial A}{\partial v} = \alpha \frac{\partial}{\partial \lambda} (V_u -[ U,V ])
+ \frac{\partial^2 M}{\partial v \partial u}
+ [ \frac{\partial M}{\partial v}, U]
+ [ M, V_u - [U, V]] + (U' \phi)_v,
$$
$$
\frac{\partial B}{\partial u} = \alpha \frac{\partial}{\partial \lambda} V_u
+ \frac{\partial^2 M}{\partial u \partial v} + [ M_u, V]
+ [M, V_u ] + (V' \phi)_u.
$$
Then substituting these into (3.16) and simplifying, we find that
$$
\frac{\partial A}{\partial v}
- \frac{\partial B}{\partial u} + [ A, V] + [ U,B] =-
[M, [U,V]]+ [ M, U],V] + [U, [M,V]] 
$$
$$
+ (U' \phi)_v
- (V' \phi)_u + [ U' \phi, V] +[ U, V' \phi ].
$$
Using Jacobi's identity, the first three terms combine to
give zero, and the last terms are the Fr\'echet derivative
of (3.4), so (3.16) holds.  $\clubsuit$

Let us apply Theorem 3.4 to the case of the sine-Gordon equation
which is given by
$$
\vartheta_{uv} = \sin \, \vartheta.
\eqno(3.38)
$$
In (3.38), $\vartheta (u,v)$ is a real, scalar function and time
is denoted by $v$. Define $U (u, v, \lambda)$ and $V (u,v, \lambda)$
in terms of the Pauli matrices as
$$
U = \frac{i}{2} (- \vartheta_u \sigma_1 + \lambda \sigma_3),
\qquad
V = \frac{i}{2 \lambda} ( \sin \vartheta \, \sigma_2
- \cos \vartheta \, \sigma_3).
\eqno(3.39)
$$
Let $\varphi$ be a solution of the equation
$$
\varphi_{uv} = \varphi \cos \vartheta,
\eqno(3.40)
$$
so $\varphi$ is considered to be a symmetry of (3.38),
and solutions of (3.40) contain the geometrical and generalized
symmetries of (3.38). For each $\varphi$, Theorem 3.4,
with $\alpha=0$, $M=0$ implies the surface constructed from
$$
A = \frac{i}{2} \frac{\partial \varphi}{\partial u} \, \sigma_1,
\qquad
B = \frac{i}{2 \lambda} \varphi ( \cos \vartheta \, \sigma_2 +
\sin \vartheta \, \sigma_3 ),
\eqno(3.41)
$$
has the immersion function given by $F = \Phi^{-1} \Phi' (\varphi)$.
Sine-Gordon equation (3.38) is an integrable equation and hence admits
infinitely many symmetries, which are referred to as generalized
symmetries.

Let $S$ be the surface generated by $U$, $V$, $A$, $B$ defined by
(3.39)-(3.41). The first and second fundamental forms,
Gaussian curvature and mean curvatures of this surface are given by
$$
I = \frac{1}{4} ( \varphi_u^2 \, du^2 + \frac{1}{\lambda^2} \varphi^2 \, dv^2),
\qquad
II = \frac{1}{2} ( \lambda \varphi_u \sin \vartheta \, du^2
+ \frac{1}{\lambda} \varphi \vartheta_v \, dv^2),
\eqno(3.42)
$$
$$
K = \frac{4 \lambda^2 \vartheta_v \sin \vartheta}{\varphi \varphi_u},
\qquad
H = \frac{2 \lambda ( \varphi_u \vartheta_v + \varphi \sin \vartheta)}
{ \varphi \varphi_u}.
\eqno(3.43)
$$

{\em Theorem 3.5.} Let $S$ be a regular surface defined by (3.42) and
(3.43) in terms of a generalized symmetry of sine-Gordon equation (3.38).
If $S$ is an oriented, compact and connected surface, then it is homeomorphic 
to a sphere.

{\bf Proof:} All compact, connected surfaces with the same Euler-Poincar\'e
characteristic are homeomorphic. For compact surfaces, the Euler-Poincar\'e
characteristic $\chi$ is given by
$$
\chi = \frac{1}{2 \pi} \int \int_{\Omega} \sqrt{ \det (g)} \, K \, du dv.
\eqno(3.44)
$$
From (3.43), the integrand can be worked out to be
$$
\sqrt{g} K = \lambda \vartheta_v \sin \vartheta =- \lambda
( \cos \vartheta )_v.
\eqno(3.45)
$$
Hence, $\chi$ is independent of the deformations $\varphi$, and
putting (3.45) into (3.44), we obtain
$$
\chi =- \frac{\lambda}{2 \pi} \int \int_{\Omega} ( \cos \vartheta)_v \, du dv.
\eqno(3.46)
$$
This implies that $\chi$ has the same value for all generalized
symmetries and hence for all sine-Gordon deformed surfaces.
It suffices to take a simple case to calculate $\chi$.
With $\varphi = \vartheta_v$, this is a sphere with $\chi=2$.
Hence, all deformed surfaces have the Euler-Poincar\'e
characteristic $\chi=2$.

\vspace{4mm}
\begin{center}
{\bf 4. DIFFERENTIAL FORMS, MOVING FRAMES AND SURFACES}
\end{center}

{\bf 4.1. Introduction to Moving Frames. }

The use of moving frames and exterior differentiation together
has become a powerful tool in differential geometry {\bf [16]}.
Suppose $f : M \rightarrow \mathbb R^N$ is an embedding of an
$m$-dimensional oriented smooth submanifold in $\mathbb R^N$.
The range of values for the indices is $1 \leq i,j,k,l \leq m$,
$m+1 \leq A,B,C,D \leq N$, $1 \leq \alpha, \beta , \gamma, \delta \leq N$.
Attach an orthogonal frame $(p; e_1, \cdots, e_N)$ to every
point in $M$ such that $e_i$ is a tangent vector of $M$ at $p$,
$e_A$ is a normal vector of $M$ at $p$ and $( e_1, \cdots, e_m)$
and $(e_1, \cdots, e_N)$ have the same orientation as a fixed
frame $(0, \delta_1, \cdots, \delta_N)$ in $\mathbb R^N$.
Suppose there is a frame field on an open neighborhood $U$
of $M$, which depends continuously and smoothly on the local
coordinates of $U$. Then we usually call such a local 
orthogonal frame a Darboux frame on the submanifold $M$.
There always exists a Darboux frame in a sufficiently
small neighborhood of every point in $M$, and the following
transformations apply
$$
e_i' = \sum_{j=1}^m \, a_{ij} e_j,
\qquad
e_A' = \sum_{B=m+1}^N \, a_{AB} e_B,
\eqno(4.1)
$$
where $a_{ij}$, $a_{AB}$ are smooth functions on $U$ such that
$( a_{ij} ) \in SO (m, \mathbb R)$, $(a_{AB}) \in SO ( N-m; \mathbb R)$.

Denote by $\omega_{\alpha}$, $\omega_{\alpha \beta}$ the
differential 1-forms obtained by pulling the relative components
of moving frames in $\mathbb R^N$ back to $U$ by $f^*$.
Obviously, these 1-forms on $U$ still satisfy the structure
equations
$$
d \omega_{\alpha} = \sum_{\gamma =1}^N \, \omega_{\beta} \wedge
\omega_{\beta \alpha},
\qquad
d \omega_{\alpha \beta} = \sum_{\gamma=1}^N \,
\omega_{\alpha \gamma} \wedge \omega_{\gamma \beta}.
\eqno(4.2)
$$
Since the origin $p$ of the Darboux frame is in $M$, and $e_i$
is a tangent vector of $M$ at $p$, we have
$$
dp = \sum_{i=1}^m \, \omega_i e_i,
\qquad
\omega_A =0,
\eqno(4.3)
$$
and the $\omega_i$, $1 \leq i \leq m$ are linearly independent
everywhere. Suppose
$$
I = dp \cdot dp = \sum_{i=1}^m \, ( \omega_i)^2,
\qquad
d A = \omega_1 \wedge \cdots \wedge \omega_m.
\eqno(4.4)
$$
These quantities are independent of the transformation of
Darboux frame, so they are defined on the whole manifold $M$.
They are referred to as the first fundamental form and the area
element of $M$. With $I$ as the Riemannian metric, the manifold
$M$ becomes a Riemannian manifold, so $M$ has a Riemannian metric
induced from $\mathbb R^N$. The equations of motion for a
Darboux frame can be written
$$
d e_i = \sum_{j=1}^m \, \omega_{ij} e_j +
\sum_{A= m+1}^N \, \omega_{i A} e_A,
\qquad
d e_B = \sum_{j=1}^m \, \omega_{Bj} e_j +
\sum_{A=m+1}^N \, \omega_{BA} e_A,
\eqno(4.5)
$$
where $\omega_{\alpha}$, $\omega_{\alpha \beta} =- \omega_{\beta \alpha}$ 
are the relative components which satisfy the structure equations
$$
d \omega_i = \sum_{j=1}^m \, \omega_j \wedge \omega_{ji},
\qquad
0 = \sum_{j=1}^m \, \omega_j \wedge \omega_{jA},
\eqno(4.6)
$$
$$
d \omega_{ij} = \sum_{k=1}^m \, \omega_{ik} \wedge \omega_{kj}
+ \sum_{A= m+1}^N \, \omega_{iA} \wedge \omega_{Aj},
$$
$$
d \omega_{iB} = \sum_{k=1}^m \, \omega_{ik} \wedge \omega_{kB}
+ \sum_{A=m+1}^N \, \omega_{iA} \wedge \omega_{AB},
\eqno(4.7)
$$
$$
d \omega_{AB} = \sum_{k=1}^m \, \omega_{Ak} \wedge \omega_{kB}
+ \sum_{C=m+1}^N \, \omega_{AC} \wedge \omega_{CB}.
$$
By the Fundamental Theorem of Riemannian Geometry, the first
formula of (4.6) and the skew-symmetry $\omega_{ij} +
\omega_{ji} =0$ together imply that $\omega_{ij}$ is the
Levi-Civita connection on the Riemannian manifold $M$,
$$
D e_j = \sum_{j=1}^m \, \omega_{ij} e_j.
\eqno(4.8)
$$
By the first formula in (4.5), we have that $D e_i$ is the
orthogonal projection of $d e_i$ on a tangent plane of $M$.

By Cartan's lemma {\bf [16]}, it follows from the second equation of
(4.6) that
$$
\omega_{jA} = \sum_{i=1}^m \, h_{Aji} \omega_i,
\qquad
h_{Aji} = h_{Aij}.
\eqno(4.9)
$$
Let us put
$$
II = \sum_{i,A} \, \omega_{i} \omega_{iA} e_A
= \sum_{A=m+1}^N \, ( \sum_{i,j=1}^m \,
h_{Aij} \omega_i \omega_j) e_A.
\eqno(4.10)
$$
Then $II$ is independent of transformation of Darboux
frame. It is a differential 2-form defined on the whole
manifold $M$, and taking values on the space of normal
vectors to $M$. It is called the second fundamental
form of the submanifold $M$.

The curvature form of the Levi-Civita connection on $M$ is
$$
\Omega_{ij} = d \omega_{ij} - \sum_{k=1}^m \, \omega_{ik}
\wedge \omega_{kj} = \frac{1}{2} \sum_{k,l=1}^m
\, R_{ijkl} \, \omega_k \wedge \omega_l,
\eqno(4.11)
$$
where $R_{ijkl}$ is the curvature tensor. From the first
formula in (4.7), we obtain
$$
R_{ijkl} = \sum_{A =m+1}^N ( h_{Ail} h_{A jk}
- h_{A ik} h_{A jl} ).
\eqno(4.12)
$$
This is the Gauss equation for the submanifold $M$.
The last two formulas in (4.7) are the Codazzi equations
from the theory of surfaces. For hypersurfaces in a
Euclidean space, the above formulas can be greatly
simplified.

The reason for the preceding introduction is to address
the following objectives. Let us show how the structure equations
for surfaces in $\mathbb R^3$ can be used to generate
integrable equations under a suitable choice of the differential
forms. This can be used to make a connection between these
equations and the theory of surfaces. This is on account of 
the fundamental theorem for hypersurfaces in $\mathbb R^{m+1}$.

{\em Proposition 4.1.} Suppose there exist two differential 2-forms
$$
I = \sum_{i=1}^m \, (\omega_i)^2,
\qquad
II = \sum_{i,j=1}^m \, h_{ij} \omega_i \omega_j,
$$
where the $\omega_i$ $(1 \leq i \leq m)$ are linearly independent
differential 1-forms depending on $m$ variables and
$h_{ij}=h_{ji}$ are functions of these $m$ variables.
Then a necessary and sufficient condition for a hypersurface to
exist in $\mathbb R^{m+1}$ with $I$ and $II$ as its first and 
second fundamental forms is: $I$ and $II$ satisfy the Gauss-Codazzi
equations (4.12) and (4.7). Moreover, any two such hypersurfaces
in $\mathbb R^{m+1}$ are related by a rigid motion.  $\clubsuit$

Now let us show how the structure equations for surfaces in
$\mathbb R^3$ can be used to generate integrable equations by
choosing the differential forms appropriately. This will allow
us to make a connection between integrable equations and the
theory of surfaces by means of Proposition 4.1. Moreover, 
it will be shown that these same partial differential
equations will result from the integrability condition 
of a particular linear system of equations {\bf [26]}. Thus, the idea
of a Lax pair has a geometrical connotation as well {\bf [27]}.
It will be seen that very many integrable equations which are of
interest in theoretical physics can be generated in this way {\bf [28]}.
It has also been shown that a system of differential forms
can reproduce the complete set of differential equations
generated by an $SO (m)$ matrix Lax pair {\bf [29]}.

Here it will be of interest to study how the fundamental
equations of surface theory can be used to reproduce the compatibility
conditions obtained from a linear system in matrix form.
The approach will be from the geometrical point of view
using the structure equations and particular choices for the
differential forms which appear in them. Concurrently, the linear matrix problem
will be worked out alongside so the equations obtained each way
can be compared. The coefficient matrices for the linear
systems of interest will be based on the Lie algebras $so (3)$
and $so (2,1)$, which are isomorphic to the Lie algebras
$su (2)$ and $sl (2, \mathbb R)$. It will be seen that a 
nonlinear partial differential equation which admits an
$SO (3)$ or $SO (2,1)$ Lax pair must be the Gauss equation
of the unit sphere in Euclidean space $\mathbb R^3$ or
Minkowski space.

{\bf 4.2. $SO (3)$ Lax Pair.}

Let us consider the $so (3)$ algebra first.
The general form for a differential equation in terms of
two independent variables and a single unknown function
$\varphi$ can be given in the form
$$
G ( \varphi, \varphi_x, \varphi_t, \varphi_{xx},
\varphi_{xt}, \varphi_{tt}, \cdots ) =0.
\eqno(4.13)
$$
In (4.13), $\varphi = \varphi (x,t)$ and $\varphi_{\alpha},
\varphi_{\alpha \beta}, \cdots$ with $\alpha, \beta \in
\{ t,x \}$ are the partial derivatives of $\varphi$ with respect 
to $x$ and $t$. For the case of an $SO (3)$ Lax pair, it
is required that there exist two three by three antisymmetric
matrices which can be expressed in the form
$$
U = \left(
\begin{array}{ccc}
0  &  u_{12}  &  u_{13}  \\
- u_{12} &  0  &  u_{23}  \\
- u_{13}  &  - u_{23}  &  0  \\
\end{array}  \right),
\qquad
V = \left(
\begin{array}{ccc}  
0  &  v_{12}  &  v_{13}  \\
- v_{12}  &  0  &  v_{23}  \\
- v_{13}  &  - v_{23}  &  0  \\
\end{array}   \right).
\eqno(4.14)
$$
such that the two linear systems
$$
\Phi_t = U \Phi,
\qquad
\Phi_x = V \Phi,
\eqno(4.15)
$$
are completely integrable when $\varphi$ satisfies (4.13).
It is said that (4.13) is a partial differential equation
admitting an $SO (3)$ Lax pair (4.14). The elements $u_{ij}$
and $v_{ij}$ which appear in (4.14) will depend on $\varphi$
and its derivatives up to a certain order. The function
$\Phi$ which appears in (4.15) can be thought of as a function
in $\mathbb R^3$ or $SO (3)$. In fact, all possible partial
differential equations of the form (4.13) which do admit such 
Lax pairs can be determined. The integrability condition for
(4.15) in terms of $U$ and $V$ is given as
$$
U_x - V_t + [ U, V] =0.
\eqno(4.16)
$$

{\em Theorem 4.1.} With respect to the components of the
matrices $U$ and $V$ defined by the matrices in (4.14),
the independent component equations of (4.16) take the form
$$
u_{12,x} - v_{12,t} + u_{23} v_{13} - u_{13} v_{23} =0,
\quad
u_{13,x} - v_{13,t} + u_{12} v_{23} - u_{23} v_{12} =0,
$$
$$
u_{23,x} - v_{23,t} + u_{13} v_{12} - u_{12} v_{13} =0.
\eqno(4.17)
$$
Equations (4.16) follow by using (4.14) in (4.16) and
working out all the operations to obtain the independent
components of the final matrix in (4.16). It may now be asked 
to what extent can the equations given in (4.17) be 
obtained from the structure equations given earlier
which govern the Darboux frame for a manifold or
immersed surface $M \subset \mathbb R^3$. The method
can be specialized to the case of a surface in
$\mathbb R^3$. Let $\{ p; e_1, e_2, e_3 \}$ be a
Darboux frame with origin $p$ in $M$. A set of
differential one-forms must be written down which depend on 
the functions $\{ u_{ij} \}$ and $\{ v_{ij} \}$.
First, the one-forms $\omega_i$ are defined to be
$$
\omega_1 = u_{12} \, dt + v_{12} \, dx,
\qquad
\omega_2 = u_{13} \, dt + v_{13} \, dx, 
\qquad
\omega_3 =0.
\eqno(4.18)
$$
The forms which specify the connection are
written as
$$
\omega_{12} = u_{23} \, dt + v_{23} \, dx,
\qquad
\omega_{13} = u_{12} \, dt + v_{12} \, dx,
\qquad
\omega_{23} = u_{13} \, dt + v_{13} \, dx.
\eqno(4.19)
$$
The forms given in (4.19) satisfy $\omega_{ij}+\omega_{ji} =0$,
hence the connection in (4.19) is Riemannian. Therefore,
it follows that
$$
dp = \omega_1 e_1 + \omega_2 e_2
= ( u_{12} \, dt + v_{12} \, dx ) e_1
+ ( u_{13} \, dt + v_{13} \, dx) e_2.
\eqno(4.20)
$$
The frame vectors $\{ e_i \}$ must satisfy the equations
$$
 d e_i = \sum_{j=1}^3 \omega_{ij} \, e_j.
\eqno(4.21)
$$

{\em Theorem 4.2.} The structure equations
$$
d \omega_1 = \omega_2 \wedge \omega_{21},
\qquad
d \omega_2 = \omega_1 \wedge \omega_{12},
\qquad
d \omega_{12} = \omega_{12} \wedge \omega_{32},
\eqno(4.22)
$$
and the differential forms given in (4.18) and
(4.19) imply the system of equations (4.17).

{\bf Proof:} From (4.18) and (4.19), it follows
that $d \omega_1 = ( u_{12,x} - v_{12,t} ) \,
dx \wedge dt$ and $\omega_2 \wedge \omega_{21}
= ( u_{13} v_{23} - v_{13} u_{23} ) \, dx \wedge dt$,
$d \omega_2 = ( u_{13,x} - v_{13,t} ) \, dx \wedge dt$
and $\omega_1 \wedge \omega_{12} = ( u_{23} v_{12}
- u_{12} v_{23} ) \, dx \wedge dt$, and finally,
$ d \omega_{12} = (u_{23,x} - v_{23,t}) \, dx \wedge dt$,
$\omega_{13} \wedge \omega_{32} = ( u_{12} v_{13} - u_{13}
v_{12}) \, dx \wedge dt$. Substituting these results 
into (4.22), it is found that system (4.17) results.
In fact, it can be seen that the two remaining structure
equations $d \omega_{13} = \omega_{12} \wedge \omega_{23}$ 
and $ d \omega_{23} = \omega_{21} \wedge \omega_{13}$
simply reproduce two of the equations already given in (4.17).
Since $\omega_1 \wedge \omega_{13} =0$ and
$\omega_2 \wedge \omega_{23} =0$, automatically it follows
that $\omega_1 \wedge \omega_{13} + \omega_2 \wedge \omega_{23} =0$. 
$\clubsuit$

Using these results for $\omega_i$ and $\omega_{ij}$, the
fundamental forms can be written down in terms of the
$u_{ij}$ and $v_{ij}$ as follows
$$
I = \omega_1^2 + \omega_2^2 
= ( u_{12}^2 + u_{13}^2 ) \, dt^2 +
2 ( u_{12} v_{12} + u_{13} v_{13} ) \, dt dx
+ ( v_{12}^2 + v_{13}^2) \, dx^2,
$$
$$
II = h_{11} \omega_1^2 + 2 h_{12} \omega_1 \omega_2
+ h_{22} \omega_2^2 = \omega_1 \omega_{13}
+ \omega_2 \omega_{23} = I,
\eqno(4.23)
$$
$$
III = \omega_{13}^2 + \omega_{23}^2 = I.
$$
Now $\omega_{13}  = h_{11} \omega_1 + h_{12} \omega_2$ and
$\omega_{23} = h_{21} \omega_1 + h_{22} \omega_2$, and since
$\omega_{13} = \omega_1$ and $\omega_{23} = \omega_2$, the
components $h_{ij}$ of $II$ must be $h_{11}= h_{22}=1$,
$h_{12}=h_{21}=0$. In this case, the two expressions for
$II$ in (4.23) exactly coincide. Using this information
about $h_{ij}$ and the definition of principle curvature,
it follows that $\kappa_1 = \kappa_2 =1$. Therefore,
every point of an associated surface is an umbilical point
of $M$. If $M$ is a connected surface on which every
point is an umbilical point, then $M$ must be a sphere or
a plane. It follows that the mean curvature and the 
Gaussian curvature have the values
$$
H = \frac{1}{2} ( h_{11} + h_{22} ) = \frac{1}{2}
( \kappa_1 + \kappa_2 ) =1,
\qquad
K = h_{11} h_{22} - h_{12}^2 = \kappa_1 \kappa_2 =1.
\eqno(4.24)
$$
The Gauss equation for the sphere can be obtained from
(4.17). Solving the first two equations for $u_{23}$ and
$v_{23}$ in (4.17), we obtain
$$
\begin{array}{c}
u_{23} = \dss \frac{1}{u_{12} v_{13} - u_{13} v_{12} }
[ ( v_{12,t} - u_{12,x})u_{12} + 
( v_{13,t} - u_{13,x}) u_{13} ],   \\
   \\
v_{23} = \dss \frac{1}{u_{12} v_{13} - u_{13} v_{12}}
[ (v_{13,t} - u_{13,x}) v_{13} +
( v_{12,t} - u_{12,x}) v_{12}].    \\
\end{array}
\eqno(4.25)
$$
Substituting $u_{23}$ and $v_{23}$ from (4.25) into
the third equation of (4.17) gives the following second
order partial differential equation
$$
( \frac{( v_{12,t} - u_{12,x})u_{12}}
{u_{12} v_{13} - u_{13} v_{12}} +
\frac{( v_{13,t} - u_{13,x}) u_{13}}
{u_{12} v_{13} - u_{13} v_{12}} )_x
- ( \frac{( v_{13,t} - u_{13,x}) v_{13}}
{u_{12} v_{13} - u_{13} v_{12}}
+ \frac{( v_{12,t} - u_{12,x}) v_{12}}
{u_{12} v_{13} - u_{13} v_{12}})_t 
$$
$$
+ u_{13} v_{12} - u_{12} v_{13} =0.
\eqno(4.26)
$$
This is an equation that is of the form (4.13).
Equation (4.26) is the Gauss equation for the sphere
$S^2$. Therefore, the nonlinear partial differential
equation (4.26) admits an $SO (3)$ Lax pair 
corresponding to an equation of the type (4.13).

It is convenient to refer to a partial differential
equation $Q$ as a subequation of another equation
$G ( \varphi, \varphi_t, \varphi_x, \cdots ) =0$ if
every solution of $Q=0$ also satisfies $G=0$. Clearly,
if $Q=0$ admits a Lax pair, then $Q=0$ must be a subequation
of each equation of (4.16). Conversely, if for given
$u_{13}$, $u_{12}$, $v_{12}$, $v_{13}$ with
$u_{12} v_{13} - u_{13} v_{12} \neq 0$, $Q=0$ is a
subequation of (4.26), then $Q=0$ admits a Lax
Pair in which $u_{23}$, $v_{23}$ are defined by (4.25).
In this sense, all possible equations admitting $SO (3)$
Lax pairs with $u_{12} v_{13} - u_{13} v_{12} \neq 0$
have been determined.

Defining the matrix
$$
M  =  \left(
\begin{array}{ccc}
u_{23}  &  u_{13}  &  u_{12}  \\
v_{23}  &  v_{13}  &  v_{12}  \\
\end{array}   \right),
\eqno(4.27)
$$
then if rank$(M) =2$, we can assume that $v_{13} u_{12}
- u_{13} v_{12} \neq 0$. When rank $(M) =1$, the second
row of (4.27) must be a multiple of the first row. In this case,
we have
$$
v_{23} =  \sigma u_{23},
\qquad
v_{13} = \sigma u_{13},
\qquad
v_{12} = \sigma u_{12}.
\eqno(4.28)
$$
Substituting (4.28) into the compatibility conditions
(4.17), the following conservation laws result
$$
u_{23,x} - ( \sigma u_{23})_t =0,
\qquad
u_{13,x} - (\sigma u_{13})_t =0,
\qquad
u_{12,x} - (\sigma u_{12})_t =0.
\eqno(4.29)
$$
Since the integrability condition (4.16) consists of
only one equation, we suppose (4.13) is the first
equation here, namely $u_{23,x} - (\sigma u_{23})_t=0$.
This is the integrability condition of the system
$$
\psi_t = u_{23} \psi,
\qquad
\psi_x = \sigma u_{23} \psi.
\eqno(4.30)
$$
In (4.30), $\psi$ is a real function and (4.30)
is a $U (1)$ Lax pair. These results can be summarized
in the form of the following Theorem.

{\em Theorem 4.3.} All nonlinear partial differential equations
admitting $SO (3)$ integrable systems can be obtained in
the following ways:

$(i)$ When rank $(M) =2$, the nonlinear equation is the
Gauss equation of $S^2 \subset \mathbb R^3$ or its
subequation and $u_{12}$, $u_{13}$, $v_{12}$, $v_{13}$ in
(4.27) are any given functions of $\varphi$ and derivatives of
$\varphi$ up to a certain order.

$(ii)$ When rank $(M) =1$, the nonlinear equation can be chosen
to be the equation of a conservation law $M_t + N_x =0$, 
where $N \neq 0$.

If $u_{12}$, $u_{13}$, $v_{12}$ and $v_{13}$ are given functions 
of $\varphi$ and derivatives of $\varphi$ up to a certain order
such that $u_{13} v_{13} - u_{12} v_{12} \neq 0$, then Theorem 4.3
gives a straightforward way of building all nonlinear partial
differential equations which admit $SO (3)$ Lax pairs. Substituting
this set of functions into (4.26), the corresponding nonlinear
equation (4.13) is obtained. Some examples in which this is done
will be presented now.

Example 1: Let $u_{13} = v_{12}=0$, $u_{12} = \cos ( \varphi /2)$,
$v_{13} = \sin (\varphi /2)$. Putting these in (4.26) gives
$$
\varphi_{tt} - \varphi_{xx} = -\sin (\varphi).
$$

Example 2: Let $u_{13} = v_{12} =0$, $u_{12} = \cosh ( \varphi /2)$,
$v_{13} = \sinh ( \varphi /2)$ in (4.26) gives the equation
$$
\varphi_{tt} + \varphi_{xx} = -\sinh ( \varphi).
$$

Example 3: Let $u_{13}= v_{12} =0$, $u_{12}  = v_{13} = e^{\varphi}$,
then the Liouville equation is obtained
$$
\varphi_{tt} + \varphi_{xx} =- e^{\varphi}.
$$

Example 4: Let $u_{13} = v_{12} =0$, $u_{12} = \varphi_t$ and
$u_{13} = \varphi^2$, then (4.26) gives
$$
(2 + \varphi^2) \, \varphi_t +
\frac{\varphi_{xxt}}{\varphi^2} - 2 \frac{\varphi_x \varphi_{xt}}{\varphi^3}=0.
$$

{\bf 4.3. Nonlinear Partial Differential Equations Admitting $SO (2,1)$ Lax Pairs.}

Consider nonlinear partial differential equations of the
form (4.13) which now admit the $SO (2,1)$ Lax pair with
structure identical to (4.15), but with matrices $U$ and
$V$ taking values in the Lie algebra $so (2,1)$. The case
in which the integrability condition for (4.15) is the Gauss
equation for $ H \subset \mathbb R^{2,1}$ will be examined.
The case of $S^{1,1} \subset \mathbb R^{2,1}$ has been examined
{\bf [27]}.

Let us consider the case in which the relevant matrices $U$ and
$V$ are given by
$$
U =  \left(
\begin{array}{ccc}
0  &  u_{12}  &  u_{13}  \\
u_{12}  &  0  &  u_{23}  \\
u_{13}  & - u_{23}  &  0  \\
\end{array}   \right),
\qquad
V =  \left(
\begin{array}{ccc}
0  &  v_{12}  &  v_{13}  \\
v_{12}  &  0  &  v_{23}  \\
v_{13}  & - v_{23} &  0  \\
\end{array}     \right).
\eqno(4.31)
$$
The compatibility condition (4.16) leads to the following Theorem.

{\em Theorem 4.4.} In terms of the components of the matrices
$U$ and $V$ defined by (4.31), the independent components of
(4.16) take the form
$$
u_{12,x} - v_{12,t} + u_{23} v_{13} - u_{13} v_{23}  =0,
\quad
u_{13,x} - v_{13,t} + u_{12} v_{23} - u_{23} v_{12}  =0,
$$
$$
u_{23,x} - v_{23,t} + u_{12} v_{13} - u_{13} v_{12} = 0.
\eqno(4.32)
$$
These same equations can be obtained directly from the
structure equations by specifying a set of differential forms.
First, the one forms $\omega_i$ are defined to be
$$
\omega_1 = u_{12} \, dt + v_{12} \, dx,
\qquad
\omega_2 = u_{13} \, dt + v_{13} \, dx,
\quad
\omega_3 =0.
\eqno(4.33)
$$
The forms which specify the connection are given by
$$
\omega_{12} = u_{23} \, dt + v_{23} \, dx,
\qquad
\omega_{13} = u_{12} \, dt + v_{12} \, dx,
\qquad
\omega_{23} = u_{13} \, dt + v_{13} \, dx,
\eqno(4.34)
$$
which satisfy $\omega_{ij} =- \omega_{ji}$ and $\omega_{23}= 
\omega_{32}$, so the
connection is quasi-Riemannian.

{\em Theorem 4.5.} For the space $H^2 \subset R^{2,1}$, the
structure equations (4.22) and the differential forms (4.33) and (4.34)
imply the system of equations (4.32).

{\bf Proof:} From (4.33) and (4.34), it follows that
$d \omega_1 = ( u_{12,x} - v_{12,t} ) \, dx \wedge dt$ and
$\omega_2 \wedge \omega_{21} = ( u_{13} v_{23} - u_{23} v_{13}) \, dx \wedge dt$,
moreover $ d \omega_2 =  ( u_{13,x} - v_{13,t}) \, dx \wedge dt$
and $\omega_1 \wedge \omega_{12} = ( u_{23} v_{12} - u_{12} v_{23} )
\, dx \wedge dt$, and finally $d \omega_{12} = ( u_{23,x} - v_{23,t})\,
dx \wedge dt$, with $\omega_{13} \wedge \omega_{32}
= (u_{13} v_{12} - u_{12} v_{13}) \, dx \wedge dt$.
Substituting these results into (4.22), the system of
equations (4.32) results. The remaining two structure equations which
go with (4.22) simply reproduce two of the equations present
in (4.32). Since both $\omega_1 \wedge \omega_{13} =0$ and
$\omega_{2} \wedge \omega_{23} =0$, it follows that
$\omega_1 \wedge \omega_{13} + \omega_2 \wedge \omega_{23} =0$. $\clubsuit$

The fundamental forms can be calculated according to (4.23),
and we have
$$
I = \omega_1^2 + \omega_2^2 =
( u_{12}^2 + u_{13}^2 ) \, dt^2 + 2 ( u_{12} v_{12} + u_{13} v_{13}) \, dx dt
+ (v_{12}^2 + v_{13}^2 ) \, dx^2.
\eqno(4.35)
$$
It is found that $h_{11}=1$, $h_{22}=1$, $h_{12}=h_{21}=0$, hence $H=1$
and $K=1$. Solving the first two equations of (4.32) for $u_{23}$ and $v_{23}$,
we obtain
$$
\begin{array}{c}
u_{23} = \dss \frac{1}{u_{12} v_{13} - u_{13} v_{12}}
[ u_{12} ( v_{12,t} - u_{12,x} )+ u_{13} (v_{13,t} - u_{13,x})],  \\
    \\
v_{23} = \dss \frac{1}{u_{12} v_{13} - u_{13} v_{12}}
[ v_{13} ( v_{13,t} - u_{13,x}) + v_{12} ( v_{12,t} - u_{12,x})].
\end{array}
\eqno(4.36)
$$
Using these results in the third equation of (4.32), the Gauss
equation of $H^2 \subset R^{2,1}$ is obtained
$$
( \frac{ (v_{12,t} - u_{12,x}) u_{12} +  ( v_{13,t} - u_{13,x} )u_{13} }
{ u_{12} v_{13} - u_{13} v_{12}} )_x
- ( \frac{v_{13} ( v_{13,t} - u_{13,x} ) + v_{12}
( v_{12,t} - u_{12,x})}{u_{12} v_{13} - u_{13} v_{12}} )_t
$$
$$
+ u_{12} v_{13} - u_{13} v_{12} =0.
\eqno(4.37)
$$
Let us summarize these results in the last Theorem of this
section.

{\em Theorem 4.6.} A nonlinear partial differential equation which
admits an $SO (2,1)$ Lax pair with $u_{12} v_{13} - u_{13} v_{12} \neq 0$ 
is equation (4.37) or a subequation. Equation (4.37) is the Gauss equation for
$H^2 \subset R^{2,1}$, and $u_{12}$, $u_{13}$, $v_{12}$
and $v_{13}$ are given functions of $\varphi$ and the partial 
derivatives of $\varphi$ up to a certain order.

Several examples of equations which are given by (4.37)
after picking the $u_{ij}$ and $v_{ij}$ will be given
to finish the Section.

Example 1: Let $u_{13} = v_{12}=0$, $u_{12} =\cos (\varphi/2)$,
$v_{13} = \sin ( \varphi/2)$, then (4.37) gives
$$
\varphi_{tt} - \varphi_{xx} = \sin ( \varphi).
$$

Example 2: Taking $u_{13}=v_{12}=0$, $u_{12} = \cosh( \varphi/2)$,
$v_{13}= \sinh (\varphi /2)$, then (4.37) gives
$$
\varphi_{tt} + \varphi_{xx} =  \sinh ( \varphi).
$$

Example 3: With $u_{13}= v_{12}=0$, $u_{12}= v_{13}= e^{\varphi}$.
we have
$$
\varphi_{tt} + \varphi_{xx} = e^{2 \varphi}.
$$

\begin{center}
{\bf 5. MAURER-CARTAN COCYCLES AND PROLONGATIONS}
\end{center}

To introduce the idea of cocycle, some definitions and theorems
will be given.  This section will see some development and
applications of the ideas in Section 3.

There are several equivalent definitions of
integrability of a set of 1-forms $\omega^1, \cdots, \omega^r$
defined on a manifold $M$ of dimension $n$ with $r <n$.

{\em Definition 5.1.} A set of $r$ linearly independent 1-forms
$\omega^1, \cdots, \omega^r$ is called completely integrable if
there exists in any neighborhood $U \subset M$ local coordinates
$x^1, \cdots, x^n$ such that
$$
\omega^i = \alpha_j^i \, dx^j,
\qquad
i=1, \cdots, r,
\eqno(5.1)
$$
where $\alpha_j^i$ are functions which are locally defined 
on $U$ such that $\det ( \alpha_j^i) \neq 0$.

As a direct consequence of this definition, we have that the
equations $\omega^1, \cdots, \omega^r$ define a local integral
manifold of dimension $n-r$. A necessary and sufficient condition
for a set of 1-forms $\omega^1, \cdots, \omega^r$ to be
completely integrable is provided by the Frobenius Theorem.

{\em Theorem 5.1. (Frobenius)} A set of 1-forms $\omega^1, \cdots, \omega^r$ 
on a manifold $M$ of dimension $n$ $(r<n)$ such that 
$\det ( \alpha_j^i ) \neq 0$ is completely integrable
if this set is closed
$$
d \omega^i = \omega^j \wedge \tau_j^i,
\eqno(5.2)
$$
where $\tau_j^i$, $1 \leq i,j \leq r$ are locally defined 1-forms.

Let $G$ be an $n$-dimensional connected Lie-group and let
$\Omega_M (G)$ be the exterior algebra of left-invariant forms,
the Maurer-Cartan forms $\Omega_{MC} (G) = \cup \Omega_{MC}^k (G)$
where $ \Omega_{MC}^k (G)$ is the collection of left-invariant 
$k$-forms on $G$.

If we take as a basis of $\Omega^1_{MC} (G)$ the 1-forms
$\omega^1, \cdots, \omega^r$, then every form
$\omega$ on $G$ is of the form
$$
\omega = \sum_{i_1 < \cdots < i_p} \, \alpha_{i_1 \cdots i_p} \,
\omega_{i_1} \wedge \cdots \wedge \omega_{i_p},
\eqno(5.3)
$$
where $\alpha_{i_1 \cdots i_r}$ are $C^{\infty}$-functions on $G$
and the $p$-form $\omega$ is left-invariant if and only if the
functions $\alpha_{i_1 \cdots i_r}$ are constant. Thus,
$\Omega_{MC} (G)$ is the exterior algebra over $\mathbb R$
generated by $\Omega_{MC}^1 (G)$. By the Frobenius Theorem,
$\Omega_{MC} (G)$ is closed in the following sense.

{\em Definition 5.2.} Let $\Gamma^1$ be a vector space of 
finite dimension $n$ on which an exterior differential
operator $d : \Gamma^1 \raro \Gamma^1 \wedge \Gamma^1$ is
given. The exterior algebra $\Gamma$ generated by $\Gamma^1$
and $d$ will be called a Maurer-Cartan algebra. On
$\Gamma$, we have that $dd \omega =0$, $d ( \omega + \eta)
= d \omega + d \eta$, $ d ( \omega \wedge \eta ) = 
d \omega \wedge \eta + (-1)^k \omega \wedge d \eta$,
where $\omega$ is a $k$-form.

Therefore, $\Omega_{MC} (G)$ is a Maurer-Cartan algebra
and will be called the Maurer-Cartan algebra of the
Lie group $G$. If $\omega^i$ is a basis in $\Gamma^i$,
then we have
$$
d \omega^i = c^i_{jk} \omega^j \wedge \omega^k,
\qquad
c_{jk}^i =- c_{kj}^i.
\eqno(5.4)
$$
The $c^i_{jk}$ are called structure constants of the
Maurer-Cartan algebra $\Gamma$ with respect to this basis.

{\em Definition 5.3.} Let $\Gamma$ be a Maurer-Cartan
algebra generated by a vector space $\Gamma^1$ of
dimension $n$ and $M$ a connected manifold also of
dimension $n$. Let $\Omega (M)$ be the set of
exterior differential forms on $M$ and $T^*_m$  the
cotangent space of $M$ at $m \in M$. Then $M$ is
called a Maurer-Cartan space if there exists a
morphism $\varphi : \Gamma \raro \Omega (M)$, 
such that at every point $m \in M$, the mapping
$\varphi |_{\Gamma^1} : \Gamma \raro T^*_m$ is
a bijection.

It follows that all co-tangent spaces of a Maurer-Cartan
space are isomorphic, and hence every connected open
sub-manifold of a Maurer-Cartan space is again
a Maurer-Cartan space.

If we have a Maurer-Cartan basis $\{ \omega^1, \cdots, \omega^n \}$ 
of $\Gamma$ with structure constants $c^i_{jk}$, then in the 
$\Gamma$-space $M$ we have, due to the morphism
$\varphi$, $n$ 1-forms $ \tau^1, \cdots, \tau^n$ on $M$
which form a basis of $T^*_m$ at every point $m$ of
$M$ such that $ d \tau^i = c^i_{jk} \, \tau^j \wedge \tau^k$.
The 1-forms $\tau^i = \varphi ( \omega^i)$, $i=1, \cdots, n$ satisfy
the same structural equations as those of the Maurer-Cartan
algebra $\Gamma$. The idea of $\Gamma$-cocycles play an important
role in the treatment of evolution equations.

{\em Definition 5.4.} Let $M$ be a manifold.
A $\Gamma$-cocycle is a morphism $\chi : \Gamma
\raro \Omega (M)$.

This need not be an injection or surjection.
A $\Gamma$ cocycle means only that we have on $M$
the set of 1-forms $\sigma^1, \cdots, \sigma^n$ such that
$\sigma^i = \chi ( \omega^i)$ which satisfy
$d \sigma^i = c_{jk}^i \, \sigma^j \wedge \sigma^k$.
These 1-forms need not necessarily be independent.
For example, the trivial $\Gamma$-cocycle is given
by $\sigma^1 = \cdots = \sigma^n =0$, and
every $\Gamma$ space $M$ has the injection
$\varphi : \Gamma \raro \Omega (M)$ as a $\Gamma$-cocycle.
These cocycles can be referred to as representative
cocycles on account of the following Theorem.

{\em Theorem 5.2.} Let $M$ be a $\Gamma$-space with representative
cocycle $\tau^1, \cdots, \tau^n$ and let $N$ be a
manifold with $\Gamma$-cocycle $\sigma^1, \cdots, \sigma^n$.
Then on $M \times N$, the system $\tilde{\mu}^1 = \tilde{\tau}^1
- \tilde{\sigma}^1=0, \cdots, \tilde{\mu}^n = \tilde{\tau}^n - \tilde{\sigma}^n =0$
is completely integrable.

{\bf Proof:} This result is actually a straightforward application
of the Frobenius Theorem and the fact that $ d \tau^i = c_{jk}^i \,
\tau^j \wedge \tau^k$ and $d \sigma^i = c_{jk}^i \, \sigma^j \wedge
\sigma^k$ with $i=1, \cdots, n$. Thus
$$
d \tilde{\mu}^i = ( c_{jk}^i \tilde{\tau}^j - c_{kj}^i
\tilde{\tau}^j ) \wedge \tilde{\mu}^k \equiv 0,
\qquad
mod ( \tilde{\mu}^k).
$$
From the fact that the forms $\{ \tau^i \}_1^n$ are independent,
it follows that the forms $\mu^i$ are independent and hence
determine a foliation of codimension $n$.

At this point, the complete $\Gamma$-space $M$ will be a
subgroup of the linear group $GL (n, \mathbb R)$, $n \in \mathbb N$
with Lie algebra $gl (n, \mathbb R)$, which is isomorphic
to the space of all $n \times n$ matrices, which will be called
$M (n, \mathbb R)$.

To find the Maurer-Cartan algebra of $GL (n, \mathbb R)$, we
consider the left-invariant forms $\omega_i^j$ as the
elements of a matrix, namely
$$
\omega = ( \omega_i^j ) = X^{-1} \, d X,
\eqno(5.5)
$$
where $X$ is the natural embedding of the group into
$\mathbb R^{n^2}$, and $X^{-1} dX$ will be called the
Maurer-Cartan form. If we define
$$
d \omega = ( d \omega_i^j ),
\eqno(5.6)
$$
and regard $\omega \wedge \omega$ as the matrix product with
exterior multiplication
$$
( \omega \wedge \omega)_i^j = \omega_i^k \wedge \omega_k^j.
\eqno(5.7)
$$
Upon differentiation, it  is easy to see that
$$
d \omega = d (X^{-1} dX )
= - X^{-1} dX X^{-1} \wedge dX
= - ( X^{-1} dX) \wedge (X^{-1} dX) =- \omega \wedge \omega.
\eqno(5.8)
$$
The Maurer-Cartan equation or algebra can be specified as
$$
d \omega + \omega \wedge \omega = 0.
\eqno(5.9)
$$
For subgroups of $GL (n, \mathbb R)$, not all $\omega_i^j$ 
will be independent, but with the natural embedding,
equation (5.9) still applies. For example, using
$SL (2, \mathbb R)= \{ X \in GL (2, \mathbb R) |
\det (X) =1 \}$, we can use the exponential mapping to
obtain the Maurer-Cartan algebra. The exponential map is
given by
$$
\exp: gl (2, \mathbb R) \raro GL (2, \mathbb R),
\qquad
A \raro e^{tA},
\quad
t \in \mathbb R.
\eqno(5.10)
$$
Since $e^{tA}$ should be an element of $SL (2, \mathbb R)$, 
it will hold that $ \det (e^{tA})=1$. Hence, using
$\det (e^{tA})= e^{tr (tA)}$, which holds for all $t$,
$tr (A) =0$. This implies that $\omega_1^1 + \omega_2^2 =0$,
so the matrix $\omega$ in (5.5) is traceless.
The Maurer-Cartan equation then takes the form,
$$
d \left(
\begin{array}{cc}
\omega_1^1  &  \omega_1^2  \\
\omega_2^1  &  -\omega_1^1  \\
\end{array}  \right)
+ \left(
\begin{array}{cc}
\omega_1^1  &  \omega_1^2  \\
\omega_2^1  & - \omega_1^1 \\
\end{array}  \right)
\wedge  \left(
\begin{array}{cc}
\omega_1^1  &  \omega_1^2  \\
\omega_2^1  & - \omega_1^1  \\
\end{array}   \right) = {\bf 0 }.
\eqno(5.11)
$$
The Maurer-Cartan algebra of the group $SL (2, \mathbb R)$ is
then given by (5.11) as
$$
d \omega_1^1 + \omega_1^2 \wedge \omega_2^1 =0,
\qquad
d \omega_1^2 + 2 \, \omega_1^1 \wedge \omega_1^2 =0,
\qquad
d \omega_2^1 + 2 \, \omega_2^1 \wedge \omega_1^1 =0.
\eqno(5.12)
$$
The Frobenius Theorem can be applied to the last pair
in (5.12) to conclude that $\omega_2^1 =0$ and $\omega_1^2 =0$ are 
integrable. Hence, $\omega_1^2 =0$ and $\omega_2^1 =0$
determine two foliations of $SL (2, \mathbb R)$. If we
put $\omega_1^2 =0$, then (5.12) becomes the Maurer-Cartan
algebra of a subgroup of $GL (2, \mathbb R)$,
$$
d \omega_1^1 =0, 
\qquad
d \omega_2^1 + 2 \, \omega_2^1 \wedge \omega_1^1 = 0.
\eqno(5.13)
$$
This subgroup will be called $BL (2, \mathbb R)$ here,
which, using the exponential mapping, consists of
matrices of the form
$$
\left(
\begin{array}{cc}
\alpha  &  0  \\
\beta  &  \dss \frac{1}{\alpha}  \\
\end{array}  \right)  \subset GL (2, \mathbb R),
\quad
\beta \in \mathbb R,
\quad
\alpha \in \mathbb R^+.
\eqno(5.14)
$$
By taking $\omega_2^1 =0$, the algebra reduces to
$d \omega_1^1 =0$. Let us now define new forms in order
to put (5.12) in a more convenient form. Introduce the forms
$\omega^1$, $\omega^2$ and $\omega^3$ such that
$$
\omega = \left(
\begin{array}{cc}
\omega_1^1  &  \omega_1^2  \\
\omega_2^1  & - \omega_1^1  \\
\end{array}   \right)
= \frac{1}{2}  \left(
\begin{array}{cc}
\omega^1  &  \omega^3 - \omega^2  \\
- \omega^3 - \omega^2  & - \omega^1  \\
\end{array}  \right).
\eqno(5.15)
$$
Therefore, the $\{ \omega^i \}$ satisfy the Maurer-Cartan
equations
$$
d \omega^1 = \omega^3 \wedge \omega^2,
\qquad
d \omega^2 = \omega^1 \wedge \omega^3,
\qquad
d \omega^3  = \omega^1 \wedge \omega^2.
\eqno(5.16)
$$
It is remarkable that in the case of linear prolongation
structures for exterior differential systems $\{ \alpha^i \}=0$,
the prolongation condition
$$
d \eta + \frac{1}{2} [ \eta, \eta ] = 0,
\eqno(5.17)
$$
determines Maurer-Cartan cocycles on $\mathbb R^2$.
Along transversal integral manifolds which are solutions of
$\{ \alpha^i =0 \}_{i=1}^l$, we have from (5.17)
$$
d \eta = d_M \, \eta^i \frac{\partial}{\partial y^i},
\qquad
\frac{1}{2} [ \eta, \eta ] = 
( \eta^j \wedge \frac{\partial \eta^i}{\partial y^j} ) \,
\frac{\partial}{\partial y^i}.
\eqno(5.18)
$$
The vector valued 1-form $\eta$ has been expressed as
$$
\eta = \eta^i \, \frac{\partial}{\partial y^i},
\qquad
\eta^i = A^i \, dx + B^i \, dt.
\eqno(5.19)
$$
Therefore, $\eta$ takes the form
$$
\eta = A^i \frac{\partial}{\partial y^i} \, dx
+ B^i \, \frac{\partial}{\partial y^i} \, dt
= ( A^i \, dx + B^i \, dt) \, \frac{\partial}{\partial y^i}.
\eqno(5.20)
$$
Now $A$ and $B$ in their turn can be written as combinations
of vertical vector fields $X_i (y)$ with $i=1,2,3$ such that
the coefficients depend on the variables in the base manifold
$M$, but not on the variables $y = ( y^1, \cdots, y^n )$ in the fibre.

These vector fields satisfy a complete Lie algebra structure
and we write
$$
\eta = \tilde{\sigma}^i X_i,
\eqno(5.21)
$$
with $\tilde{\sigma}^i$ 1-forms $\tilde{\sigma}^i_1 \, dx 
+ \tilde{\sigma}^i_2 \, dt$. The vector fields $X_i$ can be 
written as linear fields given by
$$
X_i = \alpha_{ij}^k \, y^j \, \frac{\partial}{\partial y^k},
\qquad
i=1,2,3.
\eqno(5.22)
$$
For this to match $\eta = \eta^i \frac{\partial}{\partial y^i}
= \tilde{\sigma}^i X_i$, we require that
$$
\eta^s = \tilde{\sigma}^i \alpha_{ij}^s y^j,
\qquad
\frac{1}{2} [ \eta, \eta ] = ( \eta^j \wedge \frac{\partial \eta^i}
{\partial y^j} ) \frac{\partial}{\partial y^i}.
\eqno(5.23)
$$
Therefore, we have
$$
d \eta = d \tilde{\sigma}^i \alpha_{ij}^n y^j 
\frac{\partial}{\partial y^n},
\qquad
\frac{1}{2} [ \eta, \eta ] =
( \tilde{\sigma}^i \alpha_{ij}^m y^j \wedge
\tilde{\sigma}^l \alpha_{lm}^n )
\frac{\partial}{\partial y^n}.
\eqno(5.24)
$$
Combining these, we obtain
$$
d \eta + \frac{1}{2} [ \eta, \eta ] =
( d \tilde{\sigma}^i \alpha_{ij}^k y^j + \tilde{\sigma}^i \alpha_{ij}^m y^j
\wedge \tilde{\sigma}^l \alpha_{lm}^n)
\frac{\partial}{\partial y^n} =0.
\eqno(5.25)
$$
It follows that
$$
d \tilde{\sigma}^i \, \alpha_{ij}^k + \tilde{\sigma}^i
\alpha_{ij}^m \wedge \tilde{\sigma}^l \alpha_{lm}^k = 0.
\eqno(5.26)
$$
If we simply put $\tilde{\sigma}^i \alpha_{ij}^k = \sigma_j^k$,
then equation (5.25) becomes
$$
d \sigma_j^k + \sigma_j^m \wedge \sigma_m^k = 0.
\eqno(5.27)
$$
Now if we set $\sigma = ( \sigma_j^k)$, then (5.27) is 
the Maurer-Cartan algebra that has been discussed and
must hold along the integral transversal manifolds
$\{ \alpha^i \}_{i=1}^l$. Some Maurer-Cartan cocycles
can be derived now for some equations by using their
prolongations.

First, consider the $sl (2, \mathbb R)$ prolongation of
the sine-Gordon equation. A different application of this formalism
has been considered  {\bf [30]}. Without deriving the prolongation,
we simply present here the required results,
$$
\eta = A \, dx + B \, dt,
\qquad
A = X_1 + p X_2,
\qquad
B = X_1 \cos u + X_3 \sin u,
\eqno(5.28)
$$
with $p= u_x$, and where the $X_j$ satisfy the algebra
$$
[ X_1, X_2] = X_3,
\qquad
[X_1, X_3 ] = X_2,
\qquad
[X_2, X_3] = X_1.
\eqno(5.29)
$$
Let us take the following basis for the $sl (2, \mathbb R)$ algebra
in terms of matrices
$$
\tilde{X}_1 = \frac{1}{2} \left(
\begin{array}{cc}
0  &  -1  \\
-1  &  0  \\
\end{array}  \right),
\quad
\tilde{X}_2 = \frac{1}{2}  \left(
\begin{array}{cc}
0  &  1  \\
-1  &  0  \\
\end{array}  \right),
\quad
\tilde{X}_3 = \frac{1}{2} \left(
\begin{array}{cc}
1  &  0   \\
0  &  -1  \\
\end{array}   \right),
\eqno(5.30)
$$
which satisfy the brackets (5.29). In terms of this basis, the
prolongation (5.28) can be written
$$
\tilde{A} = \frac{1}{2} \left(
\begin{array}{cc}
0  &  -1 + u_x  \\
-1 - u_x  &  0  \\
\end{array}   \right),
\quad
\tilde{B} = \frac{1}{2}  \left(
\begin{array}{cc}
\sin u  &  - \cos u  \\
- \cos u  &  - \sin u   \\
\end{array}  \right).
\eqno(5.31)
$$
Therefore, the Maurer-Cartan structure is
$$
\sigma = \frac{1}{2}  \left(
\begin{array}{cc}
0  &  -1 + u_x  \\
-1 - u_x  &  0  \\
\end{array}  \right) \, dx
+ \frac{1}{2}  \left(
\begin{array}{cc}
\sin u  &  - \cos u  \\
- \cos u  & - \sin u  \\
\end{array}  \right) \, dt.
\eqno(5.32)
$$
From this, the components of $\sigma$ are given as
$$
\sigma^1 = \sin u \, dt,
\qquad
\sigma^2 = dx + \cos u \, dt,
\qquad
\sigma^3 = u_x \, dx.
\eqno(5.33)
$$
Differentiating these, we obtain
$$
d \sigma^1 = (\cos u) u_x \, dx \wedge dt,
\qquad
d \sigma^2 =- \sin u \, u_x \, dx \wedge dt,
\qquad
d \sigma^3 = u_{xt} \, dt \wedge dx.
\eqno(5.34)
$$
Using these forms, it can be seen that the first two
equations of the Maurer-Cartan algebra hold identically,
and the third holds provided that $u$ satisfies the
sine-Gordon equation
$$
u_{xt} = \sin (u).
\eqno(5.35)
$$
There exists an $sl (2, \mathbb R)$ prolongation of the
KdV equation which is given by
$$
A = X_1 + u X_2,
\qquad
B = 2u X_1 + ( 2 u^2 -q) X_2 + pX_3,
\quad p=u_x, \quad q= u_{xx}.
\eqno(5.36)
$$
As a basis for the algebra $sl (2, \mathbb R)$, the following
matrices can be taken
$$
\tilde{X}_1 = \left(
\begin{array}{cc}
0  &  1  \\
0  &  0  \\
\end{array}  \right),
\quad
\tilde{X}_{2} =  \left(
\begin{array}{cc}
0  &  0  \\
1  &  0  \\
\end{array}  \right),
\quad
\tilde{X}_3 = \left(
\begin{array}{cc}
1  &  0  \\
0  &  -1  \\
\end{array}  \right).
\eqno(5.37)
$$
It is found from (5.37) that
$$
\sigma = \left(
\begin{array}{cc}
0  &  1  \\
u  &  0  \\
\end{array}  \right) \, dx + \left(
\begin{array}{cc}
u_x  &  2u  \\
2 u^2 - u_{xx}  &  -u_{x}  \\
\end{array}  \right)  \, dt.
\eqno(5.38)
$$
The components of $\sigma$ are then given by
$$
\sigma^1 = 2 u_x \, dt,
\qquad
\sigma^2 =- ( 1 +u) \, dx - ( 2u + 2 u^2 - u_{xx}) \, dt,
$$
$$
\sigma^3 = (1-u) \, dx + ( 2u - 2u^2 + u_{xx}) \, dt.
\eqno(5.39)
$$
Again, the first Maurer-Cartan equation holds identically,
and the last two hold provided that the function $u$ 
satisfies the KdV equation
$$
u_t - 6 u u_x + u_{xxx} =0.
\eqno(5.40)
$$
There is a link between these cocycles and B\"acklund transformations.
The following theorem implies that there exists a function
$f : \mathbb R^2 \raro SL (2, \mathbb R)$ with $f^* ( \omega^i) =\sigma^i$ 
for $i=1,2,3$ determined up to a left-multiplication by
$A \in SL ( 2 , \mathbb R)$.

{\em Theorem 5.3.} Let $M$ be a complete $\Gamma$-space with
representation cocycles $\tau^1, \cdots, \tau^n$ and let $N$
be a simply connected manifold with $\Gamma$-cocycle
$\sigma^1, \cdots, \sigma^n$. Then there exists a function
$f : N \raro M$ such that $f^* (\tau^i) = \sigma^i$ for
$i=1, \cdots, n$ which is determined up to a left-factor
$A \in Aut_{\Gamma} (M)$.

Thus, with every solution, or surface, in $\mathbb R^3$ of the
original evolution equation, there corresponds a two-dimensional
surface in the group $SL (2, \mathbb R)$ which may be parametrized
with the coordinates $(x,t) \in \mathbb R^2$. This fact also
leads to the idea of B\"acklund transformations. To develop this
idea, begin with the fact that $SL (2, \mathbb R)$ can be 
written as a product $SL (2, \mathbb R) = BL (2, \mathbb R) \cdot
SO (2)$ so that, for all $X \subset SL (2, \mathbb R)$, we have
$$
X = A \cdot B,
\qquad
A \in BL (2, \mathbb R),
\qquad
B \in SO (2).
\eqno(5.41)
$$
In terms of components, this can be expressed as,
$$
X = \left(
\begin{array}{cc}
\xi_1^1  &  \xi_1^2  \\
\xi_2^1  &  \xi^2_2  \\
\end{array}  \right) = \left(
\begin{array}{cc} 
\alpha  &  0  \\
\beta  & \dss \frac{1}{\alpha}  \\
\end{array} \right)
\left(
\begin{array}{cc}
\cos \gamma  &  \sin \gamma  \\
- \sin \gamma  & \cos \gamma  \\
\end{array}  \right) = \left(
\begin{array}{cc}
\alpha \cos \gamma  &  \alpha \sin \gamma  \\
\beta \cos \gamma - \dss \frac{1}{\alpha} \sin \gamma &
\beta \sin \gamma + \dss \frac{1}{\alpha} \cos \gamma  \\
\end{array}  \right).
\eqno(5.42)
$$
By identifying corresponding terms, (5.42) implies that
$$
\xi_1^1 = \alpha \cos \gamma,
\quad
\xi_1^2 = \alpha \sin \gamma,
\quad
\xi_2^1 = \beta \cos \gamma - \frac{1}{\alpha} \sin \gamma,
\quad
\xi_2^2 = \beta \sin \gamma + \frac{1}{\alpha} \cos \gamma.
\eqno(5.43)
$$
From the first two equations in (5.43), we obtain
$$
\alpha= \sqrt{(\xi_1^1)^2 + ( \xi_1^2)^2},
\quad
\beta \cos \gamma = \xi_2^1 +
\frac{\xi_1^2}{(\xi_1^1)^2 + (\xi_1^2)^2},
\quad
\beta \sin \gamma = \xi_2^2 - \frac{\xi_1^1}{(\xi_1^1)^2 + ( \xi_1^2)^2}.
\eqno(5.44)
$$
From the results in (5.44), we obtain $\beta$ in the form
$$
\beta = \frac{\xi_2^2}{\xi_1^2} \alpha - \frac{\xi_1^1}{\xi_1^2 \, \alpha}
= \frac{\xi_2^1}{\xi_1^1} \alpha - \frac{\xi_1^2}{\xi_1^1 \alpha},
\eqno(5.45)
$$
and also,
$$
\beta = \xi_2^1, \quad
if  \quad  \xi_1^2 =0,
\qquad
\beta= \xi_2^2, \quad  if \quad  \xi_1^1 =0.
$$
Finally, $\gamma$ is defined by the equation
$$
\cos \gamma = \frac{\xi_1^1}{\alpha},
\qquad
\sin \gamma = \frac{\xi_1^2}{\alpha}.
\eqno(5.46)
$$
With this type of decomposition for $SL (2, \mathbb R)$ as a product,
the Maurer-Cartan form can be written as
$$
\omega = X^{-1} d X = B^{-1} ( A^{-1} \, dA) \cdot B
+ B^{-1} \cdot dB.
\eqno(5.47)
$$
An interesting identification can be made on the basis of
(5.47). Here $A^{-1} dA$ can be regarded as the Maurer-Cartan form
of $BL (2, \mathbb R)$ and $B^{-1} dB$ that of $SO (2)$. For the
particular choice of $A$ and $B$ given in (5.42), these can be
calculated exactly,
$$
A^{-1} dA = \left(
\begin{array}{cc}
\dss \frac{1}{\alpha} \, d \alpha  & 0  \\
- \beta \, d \alpha + \alpha \, d \beta  &  - \dss \frac{1}{\alpha} \, d \alpha \\
\end{array}  \right),
\qquad
B^{-1} d B = \left(
\begin{array}{cc}
0  &  d \gamma  \\
- d \gamma  &  0  \\
\end{array}   \right).
\eqno(5.48)
$$
Substituting (5.48) into $\omega$ in (5.47), we obtain
$$
\omega = \left(
\begin{array}{cc}
\cos 2 \gamma \cdot \dss \frac{1}{\alpha} \, d \alpha + \frac{1}{2}
\sin 2 \gamma \cdot ( \beta \, d \alpha - \alpha \, d \beta)  &
\sin 2 \gamma \cdot \dss \frac{1}{\alpha} \, d \alpha + \sin^2 \gamma \cdot
( \beta \, d \alpha - \alpha \, d \beta )  \\
\sin 2 \gamma \cdot \dss\frac{1}{\alpha} \, d \alpha - \cos^2 \gamma (
\beta \, d \alpha - \alpha \, d \beta )  &  - \cos 2 \gamma \cdot
\dss \frac{1}{\alpha} \, d \alpha - \frac{1}{2} \sin 2 \gamma \cdot
( \beta \, d \alpha - \alpha \, d \beta )  \\
\end{array}  \right)
$$
$$
+  \left(
\begin{array}{cc}
0  &  d \gamma  \\
- d \gamma  &  0  \\
\end{array}  \right).
\eqno(5.49)
$$
Taking $\omega$ to be of the form (5.15), the $\omega^j$
can be solved for and must be given by
$$
\omega^1 = 2 \cos 2 \gamma \cdot \frac{1}{\alpha} \, d \alpha 
+ \sin 2 \gamma \cdot ( \beta \, d \alpha - \alpha \, d \beta ),
\qquad
\omega^2 =- 2 \sin 2 \gamma \cdot \frac{1}{\alpha} \, d \alpha +
\cos 2 \gamma \cdot ( \beta \, d \alpha - \alpha \, d \beta ),
$$
$$
\omega^3 = \beta \, d \alpha - \alpha \, d \beta + d ( 2 \gamma).
\eqno(5.50)
$$
This set of forms can be written in a much more compressed
form if we introduce forms $\tau^i$ defined as
$$
\tau^1 = \frac{2}{\alpha} \, d \alpha,
\qquad
\tau^2 = \beta \, d \alpha - \alpha \, d \beta,
\qquad
\psi= 2 \gamma.
\eqno(5.51)
$$
Then the set (5.50) reduces to the form
$$
\left(
\begin{array}{c} 
\omega^1  \\
\omega^2  \\
\end{array}   \right)  =  \left(
\begin{array}{cc}
\cos \psi & \sin \psi  \\
- \sin \psi  &  \cos \psi  \\
\end{array}   \right)  \left(
\begin{array}{c}
\tau^1  \\
\tau^2  \\
\end{array}  \right),
\qquad
\omega^3 = \tau^2 + d \psi.
\eqno(5.52)
$$
Now $\tau^1$ and $\tau^2$ can be thought of as
forms on $BL (2, \mathbb R)$ satisfying the
Maurer-Cartan algebra
$$
d \tau^1 =0, 
\qquad
d \tau^2 = \tau^1 \wedge \tau^2.
\eqno(5.53)
$$
There exists a function $f: \mathbb R^2 \raro SL (2, \mathbb R)$
such that $f^* ( \omega^i) = \sigma^i$, $i=1,2,3$.
Consequently, it follows from (5.53) that
$$
\left(
\begin{array}{c}
\sigma^1  \\
\sigma^2  \\
\end{array}  \right) =  \left(
\begin{array}{cc}
\cos \psi  &  \sin \psi  \\
- \sin \psi  &  \cos  \psi  \\
\end{array}   \right)  \left(
\begin{array}{c}
\tilde{\sigma}^1 \\
\tilde{\sigma}^2 \\
\end{array}  \right),
\qquad
\sigma^3 = \tilde{\sigma}^2 + d \psi,
\eqno(5.54)
$$
where $\psi = \psi(x,t)$ and $\tilde{\sigma}^i
= f^{*} ( \tau^i)$, $i=1,2$. Of course, the relations
$f^* (d \tau) = d f^* ( \tau)$ and $f^* ( \tau^1 )
\wedge f^* ( \tau^2)$ also hold. Using these, the
Maurer-Cartan algebra is transformed into
$$
d \tilde{\sigma}^1 =0,
\qquad
d \tilde{\sigma}^2 = \tilde{\sigma}^1 \wedge \tilde{\sigma}^2.
\eqno(5.55)
$$
Moreover, it follows that (5.54) can be inverted to the form
$$
\left(
\begin{array}{c}
\tilde{\sigma}^1  \\
\tilde{\sigma}^2  \\
\end{array}  \right)  =  \left(
\begin{array}{cc}
\cos \psi  & - \sin \psi  \\
\sin \psi  &  \cos \psi  \\
\end{array}  \right) \left(
\begin{array}{c}
\sigma^1  \\
\sigma^2  \\
\end{array}  \right), 
\qquad
\tilde{\sigma}^2 = \sigma^3 - d \psi.
\eqno(5.56)
$$
Eliminating $\tilde{\sigma}^2$, an expression for $d \psi$
results
$$
d \psi = \sigma^3 - \sin \psi \, \sigma^1 - \cos \psi \, \sigma^2.
\eqno(5.57)
$$
If we suppose that $u$ satisfies the sine-Gordon equation,
the cocycle of the sine-Gordon equation (5.33) can be
substituted into (5.57) to give
$$
d \psi = u_x \, dx - \sin \psi \sin u \, dt - \cos \psi \, dx
- \cos \psi \, \cos u \, dt.
\eqno(5.58)
$$
Collecting coefficients of $dx$ and $dt$, this implies by
using $d \psi = \psi_x \, dx + \psi_t \,dt$ that the $\psi$
derivatives are determined as
$$
\psi_x = u_x - \cos \psi,
\qquad
\psi_t =- \sin \psi \, \sin u - \cos \psi \cos u =- 
\cos ( \psi - u).
\eqno(5.59)
$$
This work has led to a very important result. Equation (5.59)
is an example of a B\"acklund transformation. This transforms
solutions $u$ of the sine-Gordon equation into the solutions of another 
equation. To write the other equation, $u= u(x,t)$ is eliminated
from (5.59). Differentiating the first equation in (5.59)
with respect to time, we have
$$
\psi_{xt} = u_{xt} + \sin \psi \, \psi_t = \sin u +
\sin \psi \, \psi_t,
$$
which implies that
$$
\sin u = \psi_{xt} - \sin \psi \, \psi_t.
\eqno(5.60)
$$
Substituting this into the second equation in (5.59),
we can obtain $\cos u$ as
$$
\cos u =- \tan \psi \, \psi_{xt} - \cos \psi \, \psi_t.
\eqno(5.61)
$$
Squaring (5.60) and (5.61) and then adding the results,
all dependence on $u$ goes and we are left with an
equation for the function $\psi$
$$
\psi_{xt}^2 = \cos^2 \psi \, ( 1 - \psi^2_t).
\eqno(5.62)
$$
An auto-B\"acklund transformation can also be constructed,
such that the transformed equation is again the sine-Gordon
equation. To this end, we substitute $\tilde{u} =u -2 \psi + \pi$
into (5.59) to obtain
$$
( \frac{\tilde{u} + u}{2} )_x = \sin ( \frac{\tilde{u} -u}{2}),
\qquad
( \frac{\tilde{u} -u}{2} )_t = \sin ( \frac{\tilde{u} +u}{2}).
\eqno(5.63)
$$
If $u$ is eliminated from (5.63), it is found that
$\tilde{u}$ again satisfies the sine-Gordon equation
$\tilde{u}_{xt} = \sin ( \tilde{u} )$.

\begin{center}
{\bf 6. THE GENERALIZED WEIERSTRASS SYSTEM INDUCING SURFACES OF 
CONSTANT AND NONCONSTANT MEAN CURVATURE}
\end{center}

{\bf 6.1. Generalized Weierstrass Representations.}

The theory of immersion and deformations of surfaces has been an
important part of classical differential geometry, and many methods
have been used to describe immersions and types of deformations as well.
The generalized Weierstrass representation put forward first by
Konopelchenko and Taimanov {\bf [31]} is particularly useful in
considering these particular kinds of problems, which will be of interest here.

Surfaces and their dynamics are very important ingredients in 
a great number of phenomena in physics and applied mathematics as
mentioned in the Introduction {\bf [32]}. They appear in the study of surface waves, 
shock waves, deformations of membranes, and many problems in
hydrodynamics connected with the motion of boundaries between
regions of differing densities and viscosities {\bf [33-36]}. 
Of special interest is the case of surfaces which have zero
mean curvature and such surfaces are referred to as minimal surfaces.
The most general method for constructing minimal surfaces in
three-dimensional Euclidean space was introduced by
Weierstrass, and we begin by reviewing this {\bf [37-38]}.

Let us take a pair of functions $( \psi_1, \psi_2 )$ such that 
$\psi_1$ is antiholomorphic and $\psi_2$ is holomorphic.
Let us suppose that these functions are defined in the same
simply connected domain $S$ in the complex plane {\bf [39]}. We have the system 
of equations
$$
\partial \psi_1 =0,
\qquad
\bar{\partial} \psi_2 =0.
\eqno(6.1)
$$
The bar denotes complex conjugation and the derivatives are
abbreviated $\partial = \partial / \partial z$ and
$\bar{\partial} = \partial / \partial \bar{z}$. In terms
of these functions, let us define the mapping $T$ by the
following formulas
$$
T : S \raro \mathbb R^3,
\quad
z \in S \raro ( X_1 (z, \bar{z}), X_2 (z, \bar{z}),
X_3 ( z, \bar{z})) \in \mathbb R^3,
$$
where the $X_j$ are determined by
$$
X_1 + i X_2 =i \int_{\Gamma} ( \bar{\psi}_1^2 \, dz' - 
\bar{\psi}_2^2 \, d \bar{z}' ),
\quad
X_1 - i X_2 =i \int_{\Gamma} ( \psi_2^2 \, dz' -
\psi_1^2 \, d \bar{z}'),
$$
$$
X_3 =- \int_{\Gamma} (\bar{\psi}_1 \psi_2 \, dz'
+ \psi_1 \bar{\psi}_2 \, d \bar{z}').
\eqno(6.2)
$$
The integrals are taken over any path $\Gamma$ which lies in
$S$ and connects the point $z$ to some initial point $z_0$. 
From (6.1), it follows that the integrands are closed forms
and hence the values of the integrals do not depend on the choice
of the path $\Gamma$. Weierstrass showed that the surface
$T (S)$ is minimal in the sense that its mean curvature vanishes 
everywhere.

To begin to generalize this, suppose the functions $\psi_1$ and
$\psi_2$ satisfy the more general system of equations
$$
\begin{array}{c}
\partial \psi_1 = \dss \frac{1}{2} p (z, \bar{z}) H \psi_2,  \qquad
\bar{\partial} \psi_2 =- \dss \frac{1}{2} p (z, \bar{z}) H \psi_1,  \\
   \\
    p (z, \bar{z}) = | \psi_1 (z, \bar{z}) |^2 +
|\psi_2 (z, \bar{z}) |^2.     \\
\end{array}
\eqno(6.3)
$$
and their complex conjugates, with real potential $p (z, \bar{z})$.
The integrals (6.2) then define the coordinates of a surface in three-dimensional
Euclidean space. This was first put forward by
Konopelchenko and Taimanov {\bf [29]}. The mean curvature function is
$H (z, \bar{z})$ in (6.3). It will be seen here
how (6.3) can arise. The coordinates
$(z, \bar{z})$ are conformal and in terms of these, the metric
and Gaussian curvature are given by
$$
p (z, \bar{z})^2 \, dz d \bar{z},
\qquad
K=- \frac{1}{p^2} \partial \bar{\partial} \log p.
\eqno(6.4)
$$
We can now ask how wide is the class of surfaces represented
by the Weierstrass formulas (6.3).

Let $F : \Sigma \raro \mathbb R^3$ be a regular mapping of
the domain $\Sigma$ of the complex plane with coordinates
$(z, \bar{z})$ into three-dimensional Euclidean space,
and metric tensor given by (6.4) {\bf [39]}. In this case, the vector
$$
G(z) = ( \partial F_1, \partial F_2, \partial F_3),
\eqno(6.5)
$$
satisfies the equation
$$
( \partial F_1)^2 + ( \partial F_2)^2 + ( \partial F_3)^2 = 0.
\eqno(6.6)
$$
Therefore,
$$
( F_x - i F_y, F_x - i F_y ) = (F_x, F_x)
- (F_y, F_y) =0.
$$
This immediately follows from the formula
$G (z) = \partial F = \frac{1}{2} ( F_x -i F_y)$ and the
condition that the metric is conformally Euclidean
$( F_x, F_x) = (F_y, F_y)$, $(F_x, F_y)=0$.
The subvariety $Q \subset \mathbb CP^1$ is defined
in terms of the homogeneous coordinates
$( \phi_1, \phi_2, \phi_3 )$ by
$$
\phi_1^2 + \phi_2^2 + \phi_3^2 =0.
$$
It is diffeomorphic to the Grassmann manifold $G_{3,2}$
formed by two-dimensional subspaces of $\mathbb R^3$.
This diffeomorphism is given by the mapping $G_{3,2} \raro Q$,
which assigns the point $(a_1 + i b_1, a_2 +i b_2,
a_3 + i b_3) \in Q$ to the plane generated by the pair
of unit vectors $( a_1, a_2, a_3)$ and $( b_1, b_2, b_3)$.
Thus, $G$ can be regarded as the Gauss map.
The Gauss map defined in this way for the surface (6.2) takes 
the form
$$
G (z) = ( \frac{i}{2} ( \bar{\psi}_1^2 + \psi_2^2 ),
\frac{1}{2} ( \bar{\psi}_1^2 - \psi_2^2 ), - \bar{\psi}_1 \psi_2 ).
\eqno(6.7)
$$
Solving (6.6) and (6.7) for $\psi_1^2$ and $\psi_2^2$, we obtain
$$
\psi_1^2 = \bar{\partial} F_2 + i \bar{\partial} F_1,
\qquad
\psi_2^2 =- \partial F_2 -i \partial F_1.
$$
These results give rise to the following Proposition.

{\em Proposition 6.1.} Every regular conformally Euclidean
immersion of a surface into three-dimensional Euclidean space is 
locally defined by the generalized Weierstrass formulas
(6.2)-(6.3).

{\bf Proof:} Assume that $\partial F_3 \neq 0$, otherwise
change coordinates in $\mathbb R^3$. Let us compare
$G (z) = \frac{1}{2} ( F_x -i F_y)$ with $G$ in the form
of the Gauss map and define the functions
$$
\varphi_1^2 = \bar{\partial} F_2 +i \bar{\partial} F_1,
\qquad
\varphi_2^2 =- \partial F_2 - i \partial F_1, 
\eqno(6.8)
$$
and their conjugates. In fact, these imply that
$- (\partial F_1)^2 - (\partial F_2)^2 =
\bar{\varphi}_1^2 \varphi_2^2$ and therefore,
$$
( \partial F_1)^2 + ( \partial F_2)^2 +
( \partial F_3)^2 =0.
$$
Also (6.8) can be solved for $\varphi_1$ and $\varphi_2$ as square 
roots. Recall the definition of the second fundamental form
$h_{ij}$. Let the metric tensor on the surface $ F : \Sigma
\raro \mathbb R^3$ be given by (6.4). Take an orthonormal basis
in the tangent plane at the point $z$,
$$
e_1 = \frac{1}{p} F_x,
\qquad
e_{2} = \frac{1}{p} F_y,
$$
and extend it to a basis in $\mathbb R^3$ by including a
unit normal vector
$$
e_3 = e_1 \times e_2.
$$
Components of the curvature tensor are defined by the decomposition
formulas
$$
F_{xx} = p_x e_1 - p_y e_2 + p^2 h_{11} e_3,
\quad
F_{xy} = p_y e_1 + p_x e_2 + p^2 h_{12} e_3,
\quad
F_{yy} =- p_x e_1 + p_y e_2 + p^2 h_{22} e_3.
\eqno(6.9)
$$
Given the system (6.9), we derive the associated system
satisfied by $(\varphi_1, \varphi_2)$. We make use of the
identity
$$
\partial \bar{\partial} = \frac{1}{4} ( \partial_x^2 + \partial_y^2).
$$
Differentiating $\varphi_1^2 = \bar{\partial} F_2 + i \bar{\partial} F_1$
with respect to $\partial$, we obtain
$$
2 \varphi_1 \partial \varphi_1 = \partial \bar{\partial} F_2 + i \partial
\bar{\partial} F_1 = \frac{1}{4} ( \partial_x^2 F_2 + \partial_y^2 F_2)
+ \frac{i}{4} ( \partial_x^2 F_1 + \partial_y^2 F_1).
\eqno(6.10)
$$
An explicit formula for $e_3$ is required and can be obtained by
starting with the representations
$$
e_1 = \frac{1}{p} F_x = \frac{1}{p} ( F_{1,x}, F_{2,x}, F_{3,x}),
\qquad
e_2 = \frac{1}{p} F_y = \frac{1}{p} ( F_{1,y}, F_{2,y}, F_{3,y}).
$$
Taking the cross product,
$$
e_1 \times e_2 = \frac{1}{p^2}  ( F_{2,x} F_{3,y} - F_{3,x} F_{2,y},
- (F_{1,x} F_{3,y} - F_{1,y} F_{3,x} ),
F_{1,x} F_{2,y} - F_{1,y} F_{2,y}).
$$
Thus,
$$
8 \varphi_1 \partial \varphi_1 = \frac{p_x}{p} F_{2,x} - \frac{p_y}{p} F_{2,y}
+ h_{11} (- F_{1,x} F_{3,y} + F_{1,y} F_{3,x}) - \frac{p_x}{p} F_{2,x} 
+ \frac{p_y}{p} F_{2,y} + h_{22} (-F_{1,x} F_{3,y} + F_{1,y} F_{3,x})
$$
$$
+i ( \frac{p_x}{p} F_{1,x} - \frac{p_y}{p} F_{1,y} + h_{11}
( F_{2,x} F_{3,y} - F_{3,x} F_{2,y}) - \frac{p_x}{p} F_{1,x}
+ \frac{p_y}{p} F_{1,y} + h_{22} (F_{2,x} F_{3,y} - F_{3,x} F_{2,y}))
$$
$$
= ( h_{11} + h_{22})( - F_{1,x} F_{3,y} + F_{1,y} F_{3,x})
+ i ( h_{11} + h_{22})( F_{2,x} F_{3,y} - F_{3,x} F_{2,y})
\eqno(6.11)
$$
$$
= (h_{11} + h_{22}) [ i F_{3,y} (i F_{1,x} + F_{2,x})
- F_{3,x} ( i F_{2,y} - F_{1,y})].
$$
Finally, it is required to replace the derivatives of $F_j$ in 
terms of the functions $\varphi_j$ and their complex
conjugates. To do this, we write explicitly,
$$
\varphi_1^2 = \frac{1}{2} ( F_{2,x} + i F_{2,y} + i F_{1,x}
-i F_{1,y}),
\quad
\varphi_2^2 = \frac{1}{2} (-F_{2,x} + i F_{2,y} -i F_{1,x}
- F_{1,y}),
\eqno(6.12)
$$
and their complex conjugate equations. From (6.12),
it follows that
$$
\varphi_1^2 - \varphi_2^2 = F_{2,x} +i F_{1,x},
\qquad
\varphi_1^2 + \varphi_2^2 =i F_{2,y} - F_{1,y}.
$$
Moreover,
$$
\partial F_3 = \frac{1}{2} ( \partial_x -i \partial_y) F_3 =-
\bar{\varphi}_1 \varphi_2,
\quad
\bar{\partial} F_3 = \frac{1}{2} ( \partial_x +i \partial_y) F_3
=- \varphi_1 \bar{\varphi}_2,
$$
$$
\partial_x F_3 =- ( \bar{\varphi}_1 \varphi_2 + \varphi_1
\bar{\varphi}_2),
\quad
i \partial_y F_3 = \bar{\varphi}_1 \varphi_2 - \varphi_1
\bar{\varphi}_2.
$$
Substituting these results into the final expression produced
in (6.11), we obtain
$$
8 \varphi_1 \partial \varphi_1 = ( h_{11} + h_{22})
[i F_{3,y} ( \varphi_1^2 - \varphi_2^2) - F_{3,x}
( \varphi_1^2 + \varphi_2^2 )]
$$
$$
= ( h_{11} + h_{22}) [ ( \bar{\varphi}_1 \varphi_2 - 
\varphi_1 \bar{\varphi}_2) ( \varphi_1^2 - \varphi_2^2)
+ ( \bar{\varphi}_1 \varphi_2 + \varphi_1 \bar{\varphi}_2)
( \varphi_1^2 + \varphi_2^2)]
$$
$$
=2 ( h_{11} + h_{22}) ( |\varphi_1|^2 + |\varphi_2|^2)
\varphi_1 \varphi_2.
$$
Solving for $\partial \varphi_1$ and using the definition
of mean curvature in terms of $h_{ij}$, we have the first
equation in (6.3). Similarly, we can work out
$$
8 \varphi_2 \bar{\partial} \varphi_2 =- (h_{11} + h_{22})
( - F_{1,x} F_{3,y} + F_{1,y} F_{3,x} + i F_{2,x} F_{3,y}
- i F_{3,x} F_{2,y}).
$$
The right-hand side of this result is identical except
for sign to what was worked out in the previous case, hence,
$$
2 \varphi_2 \bar{\partial} \varphi_2 =- 2 \varphi_1 \varphi_2
( h_{11} + h_{22})p.
$$
This is the second equation in (6.3). This finishes the proof.

Thus, Konopelchenko {\bf [19, 31, 37]} has established a connection
between certain classes of constant mean curvature surfaces
and the trajectories of an infinite-dimensional Hamiltonian
system of the form (6.3). He considered the nonlinear
Dirac-type system of equations in terms of two complex valued
functions $\psi_1$ and $\psi_2$ which, after
absorbing constants into the derivative variables, satisfy
the set,
$$
\begin{array}{ccc}
\partial \psi_1 = p \psi_2,  &  &  \bar{\partial} \psi_2 =-p \psi_1   \\
   &   &     \\
\bar{\partial} \bar{\psi}_1 =p \bar{\psi}_2,  &  & \partial \bar{\psi}_2 
=-p \bar{\psi}_1,   \\
   &   &      \\
   & p= |\psi_1|^2 + |\psi_2|^2.   &     \\
\end{array}
\eqno(6.13)
$$
System (6.13) has been referred to as the generalized
Weierstrass (GW) system in the literature recently {\bf [19]}. Using (6.13),
it can be verified that the following conservation laws hold
$$
\partial ( \psi_1^2) + \bar{\partial} ( \psi_2^2) =0,
\qquad
\bar{\partial} ( \bar{\psi}_1^2 ) + \partial 
(\bar{\psi}_2^2 ) =0,
\qquad
\partial ( \psi_1 \bar{\psi}_2 ) + \bar{\partial}
( \bar{\psi}_1 \psi_2) =0.
\eqno(6.14)
$$
Making use of these conserved quantities, there exist three
real-valued quantities $X_i (z, \bar{z})$ which are completely
determined by the following path integrals
$$
X_1 +i X_2 = 2i \int_{\Gamma} ( \bar{\psi}_1^2 \, dz' -
\bar{\psi}_2^2 \, d \bar{z}' ),
\qquad
X_1 -i X_2 = 2i \int_{\Gamma} ( \psi_1^2 \, dz' -
\psi_1^2 \, d \bar{z}'),
$$
$$
X_{3} =  -2 \int_{\Gamma} ( \bar{\psi}_1 \psi_2 \, dz' 
+ \psi_1 \bar{\psi}_2 \, d \bar{z}').
\eqno(6.15)
$$
On account of conservation laws (6.14), these integrals are
found to be independent of the path $\Gamma$ chosen. The
functions $X_i (z, \bar{z})$ can be treated as the
coordinates of a surface immersed in $\mathbb R^3$. 
The Gaussian curvature and the first fundamental form
of the surface are given by
$$
K =- \frac{\partial \bar{\partial} (\log p)}{p^2},
\quad
\Omega = 4 p^2 \, dz d \bar{z},
$$
in isothermic coordinates. There is also a current
which is conserved and given by
$$
J = \bar{\psi}_1 \partial \psi_2 - \psi_2 \partial \bar{\psi}_1.
\eqno(6.16)
$$
The current (6.16) satisfies $\bar{\partial} J =0$  modulo (6.13).
The integrability of system (6.13) has been examined
extensively  {\bf [40-43]} by using Cartan's theorem on
systems in involution using a set of differential forms
which are equivalent to system (6.13). A B\"acklund transformation
has also been determined for GW system (6.13) {\bf [44]}.

At this point, a correspondence between system (6.13) and the
two-dimensional nonlinear sigma model can be made. Introduce
the new variable $\rho$ which is defined in terms of the
$\psi_i$ as
$$
\rho = \frac{\psi_1}{\bar{\psi}_2}.
\eqno(6.17)
$$
Using (6.13), it can be seen that
$$
\partial \rho = \frac{\partial \psi_1}{\bar{\psi}_2}
- \frac{\psi_1}{\bar{\psi}_2^2} \partial \bar{\psi}_2
= ( 1 + |\rho|^2 ) \psi_2^2.
\eqno(6.18)
$$
Solving for $\psi_2^2$ in (6.18), using (6.17) to get
$\psi_1$, the following transformation from $\rho$ to the set of
$\psi_i$ is produced
$$
\psi_1 = \epsilon \rho \frac{(\bar{\partial} \bar{\rho})^{1/2}}
{1 + |\rho|^2},
\qquad
\psi_2 = \epsilon \frac{(\partial \rho)^{1/2}}
{1 + |\rho|^2},
\quad
\epsilon = \pm 1.
\eqno(6.19)
$$

{\em Proposition 6.2.} If $\psi_1$ and $\psi_2$ are solutions
of GW system (6.13), then the function $\rho$ defined by (6.17) 
is a solution of the second order sigma model system
$$
\partial \bar{\partial} \rho - \frac{2 \bar{\rho}}{1 +|\rho|^2}
\partial \rho \bar{\partial} \rho =0,
\qquad
\partial \bar{\partial} \bar{\rho} - \frac{2 \rho}
{1 + |\rho|^2} \partial \bar{\rho} \bar{\partial} \bar{\rho} =0.
\eqno(6.20)
$$

{\em Proposition 6.3.} If $\rho$ is a solution to sigma model
system (6.20), then the functions $\psi_1$ and $\psi_2$ defined 
in terms of $\rho$ by the expressions
$$
\psi_1 = \epsilon \rho \frac{(\bar{\partial} \bar{\rho})^{1/2}}
{1 + |\rho|^2},
\qquad
\psi_2 = \epsilon \frac{(\partial \rho)^{1/2}}
{1 + |\rho|^2},
\eqno(6.21)
$$
satisfy GW system (6.13).

Equivalently, given a solution to sigma model (6.20), a surface
can be obtained by calculating the $\psi_i$ by means of (6.21)
in Proposition 6.3 and then substituting the $\psi_i$ into (6.15) to obtain
the coordinates of the corresponding surface. This may seem involved,
but the classical symmetry group, and integrability, of system
(6.20) has been calculated explicitly {\bf [20,41]}. 
The symmetry structure of (6.20) is complicated 
enough to be able to generate a great variety of solutions $\rho$,
and by means of (6.19) to GW system (6.13) as well. Thus, the 
procedure produces useful solutions in this way which.
Once the $\psi_i$ have been calculated from (6.19), the coordinates
of a surface follow from (6.15). Another type of solution to the
sigma model system has also been discussed {\bf [20,41]}, and will be
given here as an example.

{\em Proposition 6.4.} Suppose that for each $i=1, \cdots, N$ the
complex valued functions $\rho_i$ satisfy sigma model system (6.20) 
as well as the conditions $| \rho_i |^2 =1$. Then the product of the
functions $\rho_i$
$$
\rho = \prod_{i=1}^N \, \rho_i,
\eqno(6.22)
$$
is also a solution to system (6.20).

Let us now discuss the calculation of an algebraic multi-soliton
solution of (6.13) and associated surface based on Proposition 6.4.
First, we look for a particular class of rational solutions to
(6.20) which admit simple poles at $ \bar{z} = \bar{a}_j$ given by
$$
\rho_j = \frac{z - a_{j}}{\bar{z} - \bar{a}_j},
\qquad
a_j \in \mathbb C,
\quad
j=1, \cdots, N.
\eqno(6.23)
$$
A more general class of rational solution to (6.20) admitting
simple poles by Proposition 6.4 is given by
$$
\rho = \prod_{j=1}^N \frac{z - a_j}{\bar{z} - \bar{a}_j}.
\eqno(6.24)
$$
This function satisfies $\partial \bar{\partial} \rho \neq 0$
and $| \rho |^2 =1$ as well. The first derivatives of $\rho$
are given as
$$
\partial \rho = \sum_{j=1}^N \frac{\rho}{z - a_j} = F(z) \rho,
\qquad
\bar{\partial} \rho =- \sum_{j=1}^N \frac{\rho}{\bar{z} - \bar{a}_j}
=- \bar{F} ( \bar{z}) \rho.
\eqno(6.25)
$$
Moreover, $p$ and current $J$ given by (6.15) are calculated to be
$$
p = \frac{1}{2} | \sum_{j=1}^N \frac{1}{ z - a_j} |,
\qquad
J = \frac{1}{4} ( \sum_{j=1}^N \, \frac{1}{z - a_j} )^2.
\eqno(6.26)
$$
For the case $N=1$, the functions $\psi_1$ and $\psi_2$ can be
substituted into relations (6.15) which give the coordinates $X_i$
of a surface. The corresponding constant mean curvature surface is
then given by the algebraic relation
$$
(( X_1)^2 + (X_2)^2 )^2 - ( 2 + \frac{a^2}{4} e^{2 X_3} )
( X_1^2 + X_2^2) + \frac{a^2}{2} e^{2 X_3} X_2 +1 -
\frac{a^2}{4} e^{2 X_3} =0.
\eqno(6.27)
$$
This type of inducing can be extended to higher dimensional
spaces, in particular, 4-dimensional Euclidean space and
Minkowski spaces. This was first proposed by Konopelchenko
and Landolfi {\bf [36]}. They consider a first order nonlinear
system of two-dimensional Dirac-type equations in terms of 
four complex valued functions $\psi_{\alpha}$ and $\varphi_{\alpha}$,
with $\alpha=1,2$. This system can be written as follows
$$
\partial \psi_{\alpha} =p \varphi_{\alpha},
\quad
\bar{\partial} \varphi_{\alpha} =-p \psi_{\alpha},
\quad
\alpha=1,2,
\eqno(6.28)
$$
$$
p = \sqrt{u_1 u_2},
\qquad
u_{\alpha} = |\psi_{\alpha}|^2 + |\varphi_{\alpha}|^2,
$$
as well as the complex conjugate equations of (6.28).
The system (6.28) possesses several conservation laws,
such as
$$
\partial ( \psi_{\alpha} \psi_{\beta}) + \bar{\partial}
( \varphi_{\alpha} \varphi_{\beta}) =0,
\qquad
\partial ( \psi_{\alpha} \bar{\varphi}_{\beta})
- \bar{\partial} ( \varphi_{\alpha} \bar{\psi}_{\beta}) = 0.
\eqno(6.29)
$$
As a consequence of these conserved quantities, there exist
four real-valued functions $X_i (z, \bar{z})$, $i=1, \cdots, 4$ 
which can be interpreted as the coordinates of a surface
immersed in Euclidean 4-space. The coordinates of the
position vector ${\bf X}= ( X_1, X_2, X_3, X_4 )$ of a
constant mean curvature surface in $\mathbb R^4$ are determined 
by the integrals
$$
X_1 = \frac{i}{2} \int_{\Gamma} [ ( \bar{\psi}_1 \bar{\psi}_2 +
\varphi_1 \varphi_2 ) \, dz' - ( \psi_1 \psi_2 + \bar{\varphi}_1
\bar{\varphi}_2 ) \, d \bar{z}' ],
$$
$$
X_{2} = \frac{1}{2} \int_{\Gamma} [( \bar{\psi}_1 \bar{\psi}_2 -
\varphi_1 \varphi_2 ) \, dz' + ( \psi_1 \psi_2 - \bar{\varphi}_1
\bar{\varphi}_2 ) \, d \bar{z}'],
$$
$$
X_3 =- \frac{1}{2} \int_{\Gamma} [( \bar{\psi}_1 \varphi_2 +
\bar{\psi}_2 \varphi_1 ) \, dz' + ( \psi_1 \bar{\varphi}_2 +
\psi_2 \bar{\varphi}_1 ) \, d \bar{z}' ],
\eqno(6.30)
$$
$$
X_4 = \frac{i}{2} \int_{\Gamma} [( \bar{\psi}_1 \varphi_2 - 
\bar{\psi}_2 \varphi_1 ) \, dz' - ( \psi_1 \bar{\varphi}_2
- \psi_2 \bar{\varphi}_1 ) \, d \bar{z}' ].
$$
In (6.30), $\Gamma$ is any contour in the complex plane.
The integrals depend only on the endpoints of the contour on
account of conservation laws (6.29). Many results for the 4-dimensional case
have been found and given in {\bf [43]}.

{\bf 6.2. A Physical Application Involving Nonlinear Sigma Models.}

Here is a physical example which should give the previous
considerations a physical perspective. 
Other interesting applications can be found in ${\bf [45,46]}$. Consider the
classical spin vector ${\bf S} = (S_1, S_2, S_3)$,
where each $S_j$ depends on the variable $t= x_0$ as well as two
spatial degrees of freedom $x_1$ and $x_2$. The $S_j$ are
real functions which satisfy
$$
S_3^2 + \kappa^2 ( S_1^2 + S_2^2 ) =1,
\eqno(6.31)
$$
and $\kappa^2 = \pm 1$ represents the curvature of spin
phase space. It is associated with the sphere $S^2$ 
when $\kappa^2 = 1$, or the pseudosphere when
$\kappa^2=-1$.

The Landau-Lifshitz equation describes the time evolution
of the spin vector and is given by
$$
\partial_{+} \vec{S} = \vec{S} \times \vec{\nabla}^2 \vec{S}.
\eqno(6.32)
$$
Let ${\cal S}$ be a matrix defined by
$$
{\cal S} = \left(
\begin{array}{cc}
S_3  &  \kappa \bar{S}_+  \\
\kappa S_+  &  - S_3    \\
\end{array}  \right),
\eqno(6.33)
$$
where $S_{\pm} = S_1 \pm i S_2$ and the bar denotes 
complex conjugation. In terms of the matrix ${\cal S}$,
the Landau-Lifshitz equation can be written as
$$
\partial_t {\cal S} = \frac{1}{2i} 
[ {\cal S}, \nabla^2 {\cal S} ].
\eqno(6.34)
$$
Introduce the variable $r$ defined in terms
of the $S_j$ as follows
$$
r = \frac{S_1 +i S_2}{1 + S_3}.
\eqno(6.35)
$$
The Cartesian components of the magnetization for $\kappa^2=1$ can
be shown to be
$$
S_1 = \frac{r + \bar{r}}{1 + |r|^2},
\qquad
S_2 = \frac{r - \bar{r}}{i ( 1 + |r|^2)},
\qquad
S_3 = \frac{1 - |r|^2}{1 + |r|^2}.
\eqno(6.36)
$$
It is clear that the components $S_j$ in (6.36)
satisfy the constraint (6.31) such that $\Delta {\cal S}$
and $\dot{\cal S}$ can be evaluated using (6.36).
Substituting the derivatives into the second form
of the Landau-Lifshitz equation, two independent 
equations in terms of the derivatives $\dot{r}$ and
$\dot{\bar{r}}$ are obtained. They are given explicitly by
$$
i \dot{r} =- \Delta r + 2 \bar{r} 
\frac{( \nabla r)^2}{1 + |r|^2},
\qquad
-i \dot{\bar{r}} = - \Delta \bar{r} + 2 r
\frac{(\nabla \bar{r} )^2}{1 + |r|^2}.
\eqno(6.37)
$$
Transforming the derivatives into complex form,
we obtain
$$
\frac{i}{4} \dot{r} + \partial \bar{\partial} r
= 2 \bar{r} \frac{\partial r \bar{\partial} r}
{1 + |r|^2},
\qquad
- \frac{i}{4} \dot{\bar{r}} + \bar{\partial}
\partial \bar{r} = 2 r \frac{\bar{\partial} \bar{r}
\partial \bar{r}}{1 + |r|^2}.
\eqno(6.38)
$$
The related case $\kappa^2=-1$ can be analyzed in
a similar way. The stereographic projection of the
pseudosphere onto  the $( S_1, S_2 )$ plane is used
$$
S_1 = \frac{\xi + \bar{\xi}}{1 - |\xi|^2},
\qquad
S_2 = \frac{\xi - \bar{\xi}}{i  ( 1 - |\xi|^2)},
\qquad
S_3 = \frac{1 + |\xi|^2}{1 - |\xi|^2},
\eqno(6.39)
$$
The equations of motion that follow are given as
$$
i \dot{\xi} =- \Delta \xi - 2 \bar{\xi}
\frac{(\nabla \xi)^2}{1 - |\xi|^2},
\qquad
-i \dot{\bar{\xi}} =- \Delta \bar{\xi} - 2 \xi
\frac{( \nabla \bar{\xi})^2}{1 - |\xi|^2}.
\eqno(6.40)
$$
Regarding $t$ as a time variable, then if
$t$ is held fixed, or if we consider $r$ or $\xi$,
depending on the case, to be independent of $t$,
then systems (6.38) and (6.40) reduce to exactly the
nonlinear sigma model equations of the form (6.20).

{\bf 6.3. Non-Constant Mean Curvature Surfaces.}

Consider next the case in which the mean curvature $H$
is not constant {\bf [47]}. Up to this point, constant $H$ with 
factor $1/2$ has been absorbed into the spatial coordinates.
Putting the numerical factor in the coordinates $(z, \bar{z})$,
the system of equations satisfied by the $\psi_{\alpha}$ which
determine a surface with mean curvature function $H$ become
$$
\begin{array}{ccc}
\partial \psi_1 = p H \psi_{2},  &   &  \bar{\partial} \psi_2 =-p H \psi_1, \\
      &    &     \\
\bar{\partial} \bar{\psi}_1 = p H \bar{\psi}_2,  &  &  
\partial \bar{\psi}_2 =-p H \bar{\psi}_1,     \\
      &    &     \\
    &  p= |\psi_1|^2 + |\psi_2|^2.  &     \\
\end{array}
\eqno(6.41)
$$
The function $H (z, \bar{z})$ denotes the mean curvature the
surface will have. Versions of Propositions 6.2-6.3
can be obtained starting with system (6.41).

{\em Proposition 6.5.} If $\psi_1$ and $\psi_2$ are 
solutions of the system (6.41) and $\rho$ is given by
(6.17), then $\psi_1$ and $\psi_2$ are obtained from
$\rho$ by means of the equations
$$
\psi_1 = \epsilon \rho \frac{(\bar{\partial} \bar{\rho})^{1/2}}
{H^{1/2} ( 1 + |\rho|^2)},
\qquad
\psi_2 = \epsilon \frac{(\partial \rho)^{1/2}}
{H^{1/2} ( 1 + |\rho|^2)},
\quad
\epsilon = \pm 1.
\eqno(6.42)
$$
Moreover, $\rho$ is a solution to the following second
order system,
$$
\partial \bar{\partial} \rho - \frac{2 \bar{\rho}}
{1 + |\rho|^2} \partial \rho \bar{\partial} \rho
= \bar{\partial} ( \ln H) \partial \rho,
\qquad
\partial \bar{\partial} \bar{\rho} - \frac{2 \rho}
{1 + |\rho|^2}
\bar{\partial} \bar{\rho} \partial \bar{\rho}=
\partial ( \ln H) \bar{\partial} \bar{\rho}.
\eqno(6.43)
$$
The converse of Proposition 6.5 holds as well.
Note the close similarity between (6.43) and (6.20).
Moreover, (6.41) implies that the $\psi_{\alpha}$ 
satisfy the set of conservation laws (6.14). Surfaces
can be induced by using solutions of (6.41) and then
substituting into (6.15), or alternatively,
given a solution of (6.43), it can be put in (6.42)
to obtain the $\psi_{\alpha}$, which are then used in
(6.15). Several results concerning this system and
details concerning the proofs can be found in {\bf [47]}.

{\em Proposition 6.6.} $(i)$ If $\psi_1$ and $\psi_2$
are solutions of (6.41) given in terms of $\rho$ in (6.17) by
(6.42), then $J$ defined by (6.16) in terms of the function
$\rho$ takes the form
$$
J (z, \bar{z}) =- \frac{\partial \rho \, \partial \bar{\rho}}
{H ( 1 + |\rho|^2)^2}.
$$

$(ii)$ Let $J$ be defined by (6.16),
then the quantity ${\cal J}$ defined by
$$
{\cal J} = J + \int_{\bar{z}_0}^{\bar{z}} \,
p^2 (z, \tau) \partial H ( z, \tau) \, d \tau,
\eqno(6.44)
$$
is conserved under differentiation with respect to
$\bar{z}$,
$$
\bar{\partial} {\cal J} =0.
\eqno(6.45)
$$
 
{\bf Proof:} $(i)$ Substituting $\psi_{\alpha}$ from 
(6.42) into (6.16) and differentiating using 
the product rule, we obtain
$$
J= -\bar{\rho} \frac{\partial \rho}
{2H^2 (1+ |\rho|^2)^2} \partial H + \bar{\rho} 
\frac{(\partial \rho)^{1/2}}{H(1 + |\rho|^2)}
\partial ( \frac{(\partial \rho)^{1/2}}{1 + |\rho|^2})
$$
$$
+\bar{\rho} \frac{\partial \rho}
{2 H^2 (1 + |\rho|^2)^2} \partial H -
\frac{(\partial \rho)^{1/2}}{H (1+ |\rho|^2)}
\partial ( \frac{\bar{\rho} ( \partial \rho)^{1/2}}
{ 1 + |\rho|^2})
$$
$$
= \frac{(\partial \rho)^{1/2}}{H ( 1 + |\rho|^2)}
[ \bar{\rho} \partial ( \frac{(\partial \rho)^{1/2}}{1 + |\rho|^2})
- \frac{\partial \bar{\rho} ( \partial \rho)^{1/2}}{1 + |\rho|^2}
- \bar{\rho} \partial ( \frac{(\partial \rho )^{1/2}}
{1 + |\rho|^2}  )] 
$$
$$
=- \frac{\partial \rho \, \partial \bar{\rho} }
{H ( 1 +|\rho|^2)^2}.
$$

$(ii)$ A proof can be found in {\bf [46]}.

{\em Proposition 6.7.} With $p$ defined in (6.41),
and $J$ in (6.16), then $p$ satisfies a second
order differential equation which involves $p$,
$J$ and the mean curvature function $H$.
The equation is given by
$$
\partial \bar{\partial} \ln p =
\frac{|J|}{p^2} - H^2 p^2.
\eqno(6.46)
$$
It has been shown {\bf [39]} that when $H$ is constant,
there is a connection between the time-independent
Landau-Lifshitz equation, which can be expressed as
$$
[ {\cal S}, \partial \bar{\partial} {\cal S} ] =0,
\eqno(6.47)
$$
and the two-dimensional nonlinear sigma model.
The matrix ${\cal S}$ will be referred to as the spin
matrix. In terms of the sigma model quantity $\rho$, the matrix
${\cal S}$ is given by
$$
{\cal S} = \frac{1}{1 + |\rho|^2} \left(
\begin{array}{cc}
1 - |\rho|^2  &  2 \bar{\rho}  \\
2 \rho  &  -1 + |\rho|^2  \\
\end{array}  \right).
\eqno(6.48)
$$
Define $f$ and $\bar{f}$ to be the $\rho$-dependent factors
on the left-hand side of the sigma model equations given in (6.20),
so in fact (6.20) can be written in the form $f=0$, $\bar{f}=0$. 
In terms of $f$ and $\bar{f}$, the matrix generated by (6.47)
is of the form
$$
[ {\cal S}, \partial \bar{\partial} {\cal S} ]
= \frac{4}{(1 + |\rho|^2)^2}   \left(
\begin{array}{cc}
\bar{\rho} f - \rho \bar{f}  &  \bar{\rho}^2 f - \bar{f}  \\
\rho^2 \bar{f} -f  &  \rho \bar{f} - \bar{\rho} f  \\
\end{array}   \right). 
\eqno(6.49)
$$
These results can be summarized as follows.

{\em Proposition 6.8.} If $\rho$ is a solution of the nonlinear
sigma model system (6.20), then the spin matrix ${\cal S}$
defined by (6.48) is a solution of the Landau-Lifshitz equation
(6.47). 

Proposition 6.8 and equation (6.47) can be modified to include the case
in which the mean curvature is not constant.
Define the matrices ${\cal R}$ and ${\cal H}$ as follows,
$$
{\cal R} = \frac{4}{( 1 + |\rho|^2)^2}
\left(
\begin{array}{cc}
- \bar{\rho} \, \partial \rho  &  \rho \bar{\partial} \bar{\rho}  \\
\partial \rho  &  - \rho^2 \bar{\partial} \bar{\rho}  \\
\end{array}  \right),
\qquad
{\cal H} =  \left(
\begin{array}{cc}
\bar{\partial} \ln (H)  &  \bar{\rho} \bar{\partial} \ln (H)  \\
\partial \ln (H)  &  \dss \frac{1}{\rho} \partial \ln (H)  \\
\end{array}  \right).
\eqno(6.50)
$$
The matrix ${\cal R}$ depends only on the variable $\rho$.
The following generalization of Proposition 6.8 can be formulated.

{\em Proposition 6.9.} If $\rho$ is a solution of the sigma model
equations (6.43) and the matrices ${\cal R}$ and ${\cal H}$ are defined 
in (6.50), then spin matrix ${\cal S}$ given by (6.48) is a solution
of the nonhomogeneous Landau-Lifshitz equation
$$
[ {\cal S}, \partial \bar{\partial} {\cal S} ]
+ {\cal R}{\cal H} =0,
\eqno(6.51)
$$
modulo (6.43).

\newpage
\begin{center}
{\bf 7. REFERENCES}
\end{center}

\noindent
$[1]$ M. D. Kruskal, Asymptology in numerical computation:
progress and plans on the Fermi-Pasta-Ulam problem, Proceedings
of the IBM Scientific Computing Symposium on Large-scale
problems in physics, IBM Data Processing Division, White Plains, NY,
43-62, (1965).  \\
$[2]$ N. J. Zabusky, Nonlinear lattice dynamics and energy 
sharing, J. Phys. Soc. Japan, {\bf 26}, 196-202 (1969).  \\
$[3]$ E. Fermi, J. Pasta and S. Ulam, Studies of Nonlinear Problems I,
Los Alamos Report LA 1940, 1955, reproduced in A. C. Newell (Ed.),
Nonlinear Wave Motion, AMS, Providence, RI, (1974).  \\
$[4]$ R. M. Miura, The Korteweg-de Vries equation, a survey of
results, SIAM Review, {\bf 18}, 412-459 (1976).  \\
$[5]$ P. D. Lax, Almost periodic solutions of the KdV equation,
Siam Review {\bf 18}, 351-375 (1976).  \\
$[6]$ C. Rogers and W. F. Schief, B\"acklund and Darboux
Transformations, Geometry and Modern Applications in
Soliton Theory, Cambridge Texts in Applied Math., Cambridge
University Press, 2002.   \\
$[7]$ A. Das, Integrable Models, World Scientific Notes in
Physics, Vol. 30, World Scientific, Singapore, 1989.  \\
$[8]$ M. J. Ablowitz, D. J. Kaup, A. C. Newell and H. Segur,
The inverse scattering transform-Fourier analysis for
nonlinear problems, Studies in Appl. Math., {\bf 53}, 249-334,
(1974).  \\
$[9]$ S. S. Chern, Surface Theory with Darboux and Bianchi,
Miscellanea Mathematica, 59-69, Springer, Berlin (1991).   \\
$[10]$ A. I. Bobenko, ``Surfaces in terms of $2 \times 2$
matrices, old and new integrable cases'', in Harmonic Maps
and Integrable Systems, Eds. A. P. Fordy and J. C. Wood,
Aspects of Mathematics (Friedr. Vieweg and Sohn, 1994),
pp. 83-127.   \\
$[11]$ H. Hopf, Differential Geometry in the Large, Lecture
Notes Math., Vol. 1000, Berlin, Heidelberg, NY, Springer, 1983.   \\
$[12]$ F. B. Estabrook, Moving frames and prolongation
algebras, J. Math. Phys. {\bf 23}, 2071-2076 (1982).  \\
$[13]$ H. D. Wahlquist and F. B. Estabrook, Prolongation Structures
of Nonlinear Evolution Equations II, J. Math. Phys.,
{\bf 17}, 1293-1297 (1976).  \\
$[14]$ N. Kamran and K. Tenenblat, On differential equations
describing pseudo-spherical surfaces, J. Diff. Equations,
{\bf 115}, 75-98 (1995).  \\
$[15]$ S. S. Chern and K. Tenenblatt, Pseudospherical
surfaces and evolution equations, Stud. Appl. Math.,
{\bf 74}, 55-83 (1986).  \\
$[16]$ S. S. Chern, W. H. Chen and K. S. Lam, Lectures
on Differential Geometry, World Scientific, Singapore,
(1999).  \\
$[17]$ D. G. Gross, C. N. Pope and S. Weinberg, Two-dimensional
Quantum Gravity and Random Surfaces, World Scientific,
Singapore, 1992.   \\
$[18]$ D. Nelson, T. Piran and S. Weinberg, Statistical 
Mechanics of Membranes and Surfaces, World Scientific,
Singapore, 1992.  \\
$[19]$ B. G. Konopelchenko, Induced Surfaces and their Integrable
Dynamics, Stud. Appl. Math., {\bf 96}, 9-51 (1996).  \\
$[20]$ P. Bracken and A. M. Grundland, Symmetry Properties and Explicit
Solutions of the Generalized Weierstrass System, 
J. Math. Phys., {\bf 42}, 1250-1282 (2001). \\
$[21]$ \"O. Ceyhan, A. S. Fokas and M. G\"urses, J. Math. Physics,
Deformations of surfaces associated with integrable Gauss-Mainardi-Codazzi
equations, {\bf 41}, 2251-2270 (2000).  \\
$[22]$ A. S. Fokas and I. M. Gelfand, Commun. Math. Phys.,
Surfaces on Lie Groups, on Lie Algebras and Their Integrability,
{\bf 177}, 203-220 (1996).  \\
$[23]$ A. S. Fokas, I. M. Gelfand, F. Finkel and Q. Liu,
Selecta Math., New Series, {\bf 6}, 347-375 (2000).  \\
$[24]$ H. Abbaspour and M. Moskowitz, Basic Lie Theory,
World Scientific, Singapore, (2007).  \\
$[25]$ P. Olver, Applications of Lie Groups to Differential
Equations, 2nd. ed., Springer, (1993).   \\
$[26]$ G. Chaohao, Soliton Theory and Its Applications,
Springer-Verlag, Berlin (1995).   \\
$[27]$ P. Bracken, Partial Differential Equations Which Admit
Integrable Systems, Int. J. of Pure and Applied Math., {\bf 43},
408-421 (2008).  \\
$[28]$ P. Bracken, Symmetry Properties of a Generalized 
Korteweg-de Vries Equation and Some Explicit Solutions'',
Int. J. of Mathamatics and Mathematical Sciences, {\bf 2005,13} 2159-2173 (2005).  \\
$[29]$ P. Bracken, Integrable Systems Determined by Differential
Forms for Moving Frames on Immersed Submanifolds, Int. J. Geometric Methods
in Modern Physics, {\bf 5}, 1041-1049 (2008).   \\
$[30]$ P. Bracken, An Exterior Differential System for a Generalized
Korteweg-de Vries Equation and its Associated Integrability, Acta
Applicandae Math., {\bf 95}, 223-231 (2007).  \\
$[31]$ B. G. Konopelchenko and I. Taimanov, Constant mean curvature surfaces via 
an integrable dynamical system, J. Phys. {\bf A 29}, 1261-1265 (1996).  \\
$[32]$ E. Radu and M. Volkov, Stationary ring solitons in field theory-Knots
and vortons, Physics Reports, {\bf 468}, 101-151 (2008).  \\
$[33]$ B. G. Konopelchenko and G. Landolfi, On rigid string instantons
in four dimensions,  Phys. Lett. {\bf B 459},
522-526 (1999).  \\
$[34]$ B. G. Konopelchenko, On solutions of the shape equation for
membranes and strings, Phys. Lett. {\bf B 414}, 58-64 (1997).  \\
$[35]$ B. G. Konopelchenko and G. Landolfi, Quantum effects for extrinsic
geometry of strings via the generalized Weierstrass representation, 
Phys. Lett., {\bf B 444}, 299-308 (1998).   \\
$[36]$ P. Bracken, A Surface Model for Classical Strings in Minkowski
Space, Phys. Lett., {\bf B 541}, 166-170 (2002).   \\
$[37]$ B. G. Konopelchenko and G. Landolfi, Generalized Weierstrass representation
for surfaces in multi-dimensional Riemann spaces, J. Geom. and Phys. {\bf 29},
319-333 (1999).  \\
$[38]$ K. Weierstrass, Fortsetzung der Untersuchung \"uber die
Minimalfl\"acher, Mathematische Werke, Vol. 3 (Verlagsbuchhandlung,
Hillesheim), 219-248, 1866.  \\
$[39]$ I. Taimanov, Modified Novikov-Veselov Equation and Differential
Geometry of Surfaces, Translations of Amer. Math. Soc., Ser. 2, {\bf 179}, 
133-150 (1997).   \\
$[40]$ P. Bracken, A. M. Grundland and L. Martina, The Weierstrass-Enneper 
System for Constant Mean Curvature Surfaces and the Completely 
Integrable Sigma Model, J. Math. Phys. {\bf 40}, 3379-3403 (1999).  \\
$[41]$ P. Bracken and A. M. Grundland, On Certain Classes of Solutions of the 
Weierstrass-Enneper System Inducing Constant Mean Curvature Surfaces,
J. Nonlinear Math. Phys.  {\bf 42}, 1250-1282 (2001).  \\
$[42]$ P. Bracken and A. M. Grundland, On the Complete Integrability of the Generalized
Weierstrass System, J. Nonlinear Math. Phys., {\bf 9}, 229-247 (2002).  \\
$[43]$ P. Bracken and A. M. Grundland, Solutions of the Generalized Weierstrass
Representation in Four-Dimensional Euclidean Space, J. Nonlinear Math. Phys.
{\bf 9}, 357-381 (2002).  \\
$[44]$ P. Bracken and A. M. Grundland, On the B\"acklund Trasformation and
the Generalized Weierstrass System, Inverse Problems, {\bf 16}, 145-153 (2000).  \\
$[45]$ P. Bracken, P. Goldstein and A. M. Grundland, On Vortex
Solutions and Links between the Weierstrass System and the Complex 
Sine-Gordon Equations, J. Nonlinear Math. Phys., {\bf 10}, 464-486 (2003).  \\
$[46]$ P. Bracken, The generalized Weierstrass system inducing
surfaces of constant and nonconstant mean curvature in Euclidean
three space, J. Comp. Appl. Math., {\bf 202}, 122-132 (2007).  \\
$[47]$ P. Bracken, The Generalized Weierstrass System for Nonconstant Mean
Curvature Surfaces and the Nonlinear Sigma Model, Acta Applicandae
Math., {\bf 92}, 63-76 (2006).  \\
\end{document}